\documentclass[12pt,preprint]{aastex}
\slugcomment{Version: \today}

\shorttitle{Optical Monitoring of V1647 Orionis}
\shortauthors{}

\begin{document}

\title{V1647~ORIONIS: OPTICAL PHOTOMETRIC AND SPECTROSCOPIC MONITORING THROUGH THE 2003--2006 OUTBURST}

\author{Colin Aspin \& Bo Reipurth}

\affil{Institute for Astronomy, University of Hawaii,\\
       640 N. A'ohoku Place, Hilo, HI 96720 \\
       {\it caa@ifa.hawaii.edu, reipurth@ifa.hawaii.edu}}

\begin{abstract} 

  We present results from an optical imaging and spectroscopic
  monitoring campaign on the young, low-mass eruptive variable star
  V1647~Orionis.  The star and associated nebulosity (McNeil's Nebula)
  were observed over the period February 2004 to February 2006 with
  observations commencing a few months after the original outburst
  event occurred.  Using the Gemini North telescope, we obtained
  multi-band optical imaging photometry and medium-resolution
  long-slit spectroscopy of V1647~Ori on an approximately monthly
  interval.  During this period, V1647~Ori remained at, or close to,
  peak brightness and then faded by 5 magnitudes to close to its
  pre-outburst brightness.  This implies an outburst timescale of
  around 27 months.  Spectral features seen in both emission and
  absorption varied considerably during the monitoring period.  For
  example, the H$\alpha$ line changed significantly in both intensity
  and profile.  We present and discuss the observed photometric and
  spectroscopic changes and consider how this eruptive event relates
  to the early formative stages of low-mass stars.

\end{abstract}


\section{INTRODUCTION}
When the amateur astronomer Jay McNeil discovered a new nebula in the L1630 molecular cloud in Orion in January 2004 (McNeil 2004), little did he know that it would spark significant and extensive worldwide follow-up investigations. The nebula he discovered has subsequently been designated "McNeil's Nebula" and its illuminating/exciting star named V1647~Orionis (Samus 2004).  To date, around 30 publications have resulted from these studies spanning the electromagnetic spectrum from X-ray to radio wavelengths.  In these research papers, considerable speculation has been put forth as to the nature of this eruptive event, yet it is generally agreed that it is the result of a rapid and massive increase in accretion onto the surface of the young star.  The reader is referred to the papers by \`Abrah\`am et al. (2004a,b, 2006), Acosta-Pulido et al. (2007), Andrews, Rothberg, \& Simon (2004), Aspin et al. (2006), Aspin, Beck, \& Reipurth (2008, herein ABR08), Aspin, Greene, \& Reipurth (2009), Aspin et al. (2009), Brice\~no et al. (2004), Brittain et al. (2007), Fedele et al. (2007a,b), Gibb et al. (2006), Grosso et al. (2005), Kastner et al. (2004, 2006), K\'osp\'al et al. (2005), Kun (2008), McGehee et al. (2004), Mosoni et al. (2005), Muzerolle et al. (2005), Ojha et al. (2004, 2006), Reipurth \& Aspin (2004), Rettig et al. (2005), Semkov (2004, 2006), Tsukagoshi et al. (2005), Vacca, Cushing, \& Simon (2004), Vig et al. (2006), and Walter et al. (2004) for further discussions.

Such eruptive events have been observed in a number of young stars dating back to the observations and discussions by Herbig (1966).  In this and subsequent papers, Herbig (1966, 1977, 1989) suggested that these eruptions can be classed as either short-term EXors events (after the progenitor EX~Lupi) lasting between a few weeks to a few years, or long-term FUors events (after the progenitor FU~Ori) lasting decades to possibly even a century.  Much emphasis has been placed on determining what type of outburst has occurred in the case of V1647~Ori, yet more compelling perhaps is understanding whether the FUor and EXor classes are fundamentally similar in their origins or whether they are the result of distinct phenomena occurring on different timescales.

If the outburst of V1647~Ori is a FUor type eruption, then it is the first in almost 40 years since that of V1057~Cyg (Welin 1971).  A large-scale effort to monitor V1647~Ori was therefore undertaken involving use of both Gemini Observatory 8--meter telescopes (North and South) and the NASA IRTF 3--meter telescope over three observing semesters and utilizing six different facility instruments. This campaign resulted in approximately monthly optical imaging and spectroscopic and near-IR spectroscopic observations of V1647~Ori.  In this paper, the first in a series describing these data, we present optical observations spanning the outburst phase, from the first follow-up observations taken soon after its discovery (taken in 2004 February and published in Reipurth \& Aspin 2004) to the time when the source had faded to its pre-outburst brightness (in February 2006). Clearly, the approximately two year lifetime of this event suggests that, at least superficially, the eruption of V1647~Ori is more similar to EXor event than those occurring in FUors.

In this paper we present optical imaging and spectroscopic observations of V1647~Ori taken between February 2004 and February 2006 and discuss the variations present in these data.  The observations and data reduction are described in $\S $2.  In $\S $3 we present details of the temporal changes observed, and in $\S $4 we consider what these observations imply regarding two physical components of the outburst, i.e. the fast wind and the accretion process.  Finally, in $\S $5 we piece together a time-line for the event and detail the evolution of the eruption as implied by the data from our monitoring campaign.

\section{OBSERVATIONS \& DATA REDUCTION}
All but one dataset presented below were acquired using the ``Frederick C. Gillett'' Gemini North telescope located on Mauna Kea, Hawaii using the facility optical imager and spectrograph, GMOS-N (Davies et al. 1997; Hook et al. 2004).  The imaging observations used standard GMOS-N g', r', i', and z' filters which were designed to be as close as possible in characteristics to the Sloan Digital Sky Survey (SDSS) filters (Fukugita et al. 1996).  For the spectroscopic observations, all but one used the blue 600 lines/mm grating and 0$\farcs$5 wide long-slit.  The one exception, the first observation taken, used the red 831 lines/mm grating, again with a 0$\farcs$5 wide long-slit.  The spectral resolution of data taken were therefore $\sim$4400 and $\sim$1700, corresponding to 0.34 and 0.45~\AA~pixel$^{-1}$ respectively.  The complete observation log is presented in Table~\ref{obslog}.  In total, we acquired photometry at 16 different epochs and spectroscopy at 15 different epochs over the eruption period.

The additional spectrum of V1647~Ori was acquired using the W.M.~Keck~II telescope located on Mauna Kea, Hawaii, using the facility high-resolution echelle spectrograph HIRES (Vogt et al. 1994).  The observations were made on UT 2004 September 24 when the star had an optical V magnitude of about 17.  The spectra covered the wavelength range 5800 to 7150~\AA\ and 7320 to 8700~\AA\ at a nominal spectral resolution of R$\sim$46,000 using a 0$\farcs$86 wide slit.  The on-source exposure time used was 60 minutes.

\subsection{Imaging data reduction and calibration}
All datasets were reduced using the Gemini IRAF data reduction package v1.8. Specifically, we used the routines {\tt gireduce} for basic instrument signature removal (trim the images, subtract the master bias image, and divide by the normalized master twilight flat field), and {\tt gmosaic} for combining the data from the three GMOS-N CCDs into one image.  In addition, multiple exposures of the region were coadded using the IRAF routine {\tt imcombine}.  Aperture photometry of V1647~Ori was performed using the Starlink program {\tt gaia} (Draper et al. 2008). An aperture diameter of 2$''$ was used for the photometric calculation and, due to the bright nebulosity immediately surrounding the star, the sky signal was estimated from numerous apertures located in blank sky regions throughout the image.  The full width half maximum (FHWM) seeing values in the acquired images are also given in Table~\ref{obslog}.

\begin{figure*}[tb] 
%\includegraphics*[angle=0,scale=0.75]{v1647ori-star-ids.ps} 
%\epsscale{1.75}
\epsscale{0.8}
\plotone{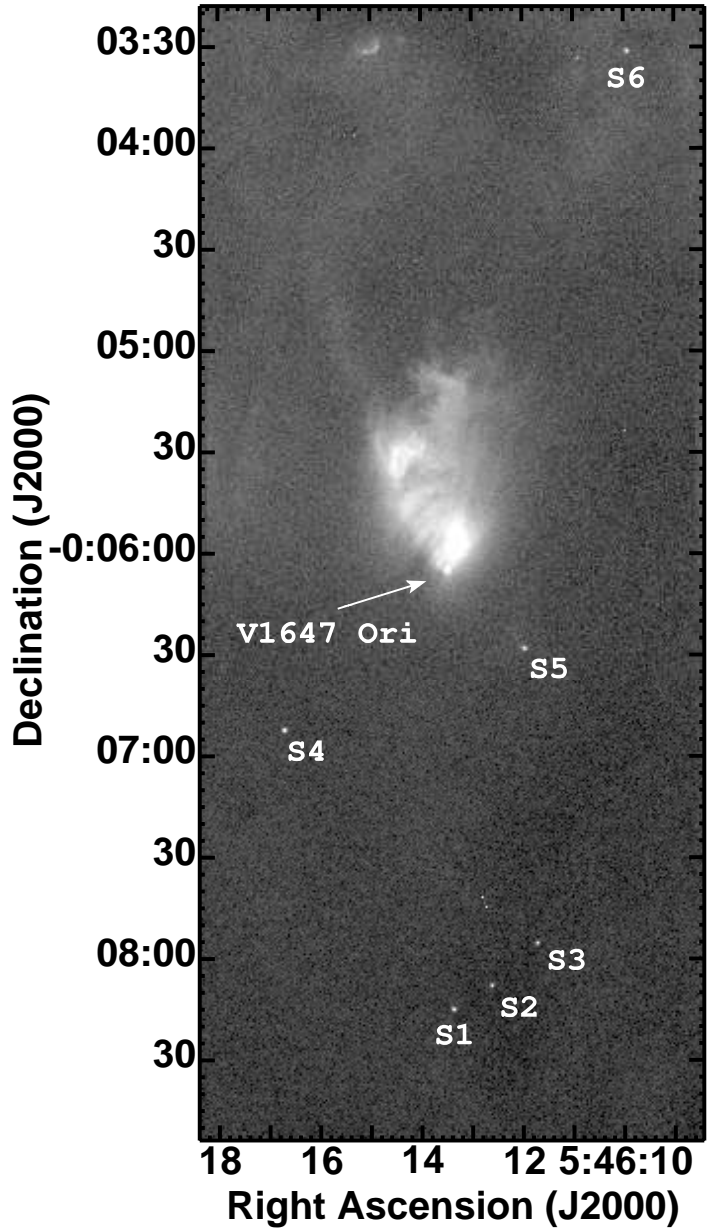} 
\caption{The region of L1630 including V1647~Ori and McNeil's Nebula.  
This r' band image was taken on Gemini North using GMOS on UT 2004 February 
14.  V1647~Ori and the six calibration sequence stars, labeled S1 to S6, are 
identified. North is at the top, East to the left.  The
scale of the image is 0$\farcs$144/pixel and the exposure time was 40 
seconds in seeing of FWHM 0$\farcs$52.
\label{idplot}}
\end{figure*}

Since some of the data were acquired during non-photometric conditions, a photometric calibration was achieved relative to a dataset taken on a photometric night, UT 2004 February 14.  A series of six field stars were used to provide a calibration sequence; their position and magnitudes are listed in Table~\ref{calstars}.  Since the region containing V1647~Ori is one with active star formation and young stars are well-known to be variable, using six field stars as calibrators was considered an acceptable way of minimizing errors introduced by intrinsic variability of the calibrator stars themselves.  These stars, together with V1647~Ori, are identified in Fig.~\ref{idplot}.  Using instrumental magnitudes derived from calibrators, we additionally studied the variability of individual calibrators themselves and concluded that uncertainties resulting from source variability were small and within the quoted uncertainties on the V1647~Ori photometry. In addition to this, and to provide a consistency check, we have used the SDSS photometry of V1647~Ori from UT Nov 11, 1998. These data were part of the early "Orion" release (Finkbeiner et al. 2004) downloaded from Princeton University\footnotemark\footnotetext{see URL
  http://photo.astro.princeton.edu/oriondatarelease}, and quoted in McGehee et al. (2004).  Photometry of the same calibration sequence stars was extracted from the SDSS data and found to be consistent with the photometry from our GMOS images within the associated errors (shown in Table~\ref{calstars}).

\subsection{Spectroscopic data reduction} 
Again, all the Gemini/GMOS data were reduced using the Gemini IRAF
data reduction package v1.8.  Specifically, we used the routines {\tt
  gsreduce}, {\tt gstransform}, {\tt gsskysub}, {\tt gscrrej}, and {\tt gsextract} for basic instrument signature removal (trim the images, subtract the master bias image, and divide by the normalized flat field), wavelength calibration, sky line subtraction, cosmic ray removal, and point-source spectrum extraction, respectively.

The resultant spectra were not flux calibrated nor were atmospheric features removed since we were primarily interested in the spectral features present and their intrinsic variability.  In addition, H$\alpha$ has been shown to be a relatively insensitive diagnostic for accretion rate due to it, and other Balmer lines, becoming optically thick at high accretion rates (Muzerolle et al. 1998), thus negating any requirement for flux determinations.  Fig.~8 of Muzerolle et al. (1998) show model predictions for H$\alpha$ flux vs. accretion rate and it is clear that the relationship saturates at around 10$^{-8}$~M$_{\odot}$~yr$^{-1}$.  Such rather low accretion rates are typically found in 'weak-line' T Tauri stars (WTTS) and low activity 'classical' T~Tauri stars (CTTS) like DN and DQ~Tau and not eruptive variables such as V1647~Ori.

For our HIRES data, a standard reduction was performed, including bias correction, flat-fielding, scattered-light correction, order extraction, and wavelength calibrations using standard routines in the IRAF echelle package.

\section{RESULTS}
\subsection{Optical photometry}
\begin{figure}[tb] 
\epsscale{1.0}
\plotone{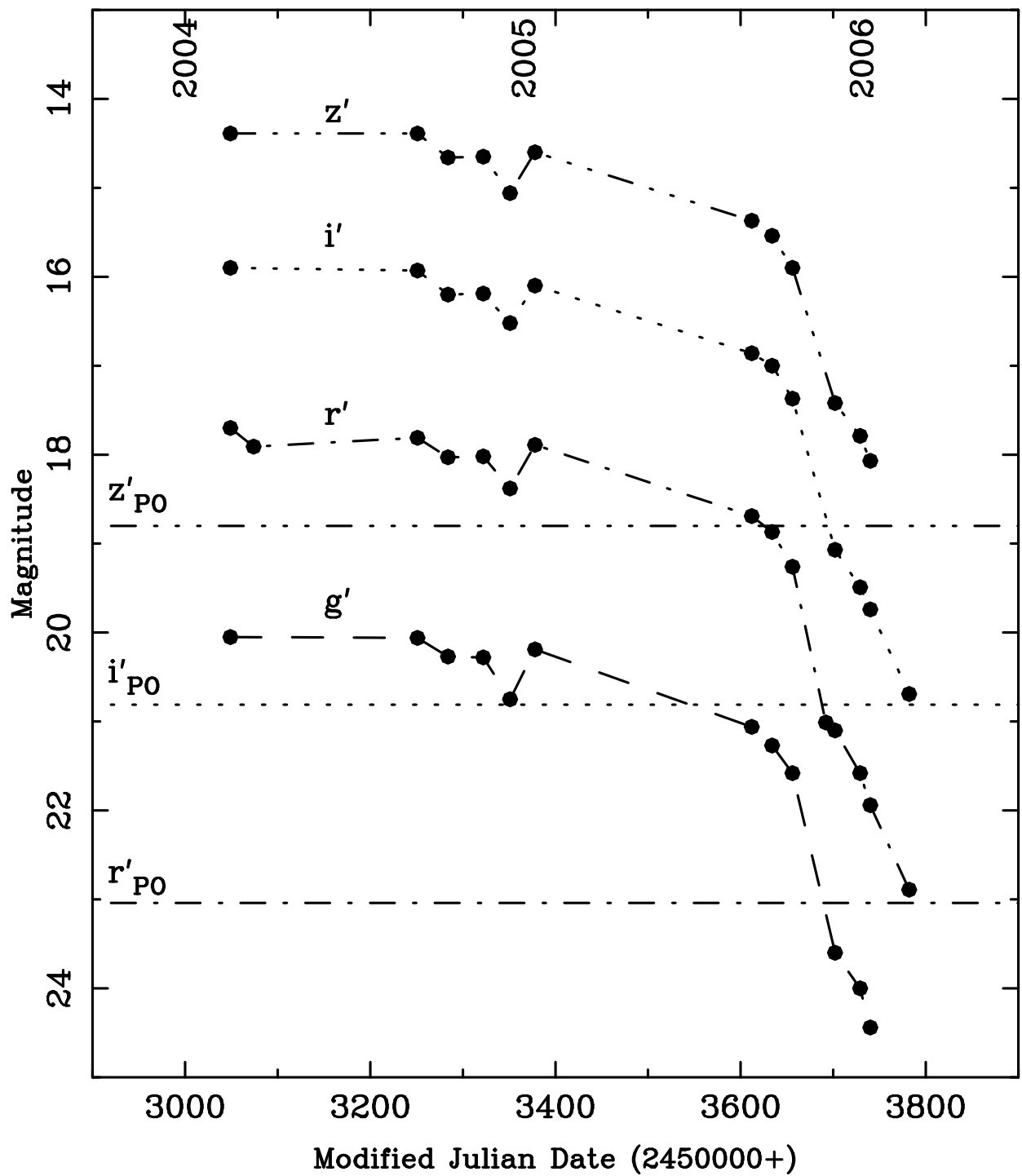} 
\caption{
Optical photometry of V1647~Ori spanning the period 2004 February to
2006 February.  The data were obtained using the Gemini facility
imager/spectrograph GMOS-N in SDSS filters g', r', i', and z'.  The
horizontal lines are the
pre-outburst brightness of V1647~Ori from the SDSS Orion survey data
and additionally presented in McGehee et al. (2004).
\label{photplot}}
\end{figure}

In Fig.~\ref{photplot} we show our optical photometry of V1647~Ori from 2004 February to 2006 February for all four passbands, namely g', r', i', and z'.  The horizontal dashed and dotted lines represent the pre-outburst SDSS brightness of the star (labeled filter$_{PO}$).  The solid lines simply join the discrete observation points as a guide to the eye.  The two long `gaps' in the observation sampling are the 2004 and 2005 summer months when Orion was not visible during the nighttime period.

Over the summer of 2004, the source appeared to remain at approximately the same brightness, however, it is likely that some small-scale variability occurred.  Our data from the 2004--2005 winter months suggests that variability at the $\sim$0.5 magnitude level was present.  Such fluctuations were also reported by McGehee et al. (2004), Semkov (2004), Walter et al. (2004), Ojha et al. (2006), Acosta-Pulido et al. (2007), and Fedele et al. (2007) during several different observing periods.

Over the summer of 2005, V1647~Ori exhibited a small yet significant decline in brightness, approximately the same in all filters, and amounting to $\le$1 magnitude.  Immediately after this period, in 2005 September, a major decline phase began.  From 2005 September onwards, the trend was for monotonic dimming, with possibly some slowing occurring in late 2005.  By the time of our last g' band photometric observation (in early 2006 January), V1647~Ori had already faded by about 4.4 magnitudes from its maximum brightness.  At r', we obtained one further observation (in 2006 mid-February) which shows a total fading of over 5 magnitudes. At this time V1647~Ori was already close to its pre-outburst SDSS r' magnitude of 23.04.

In late 2006 December we took a short R-band exposure of V1647~Ori at the University of Hawaii 2.2m telescope on Mauna Kea, Hawaii which showed that the source had remained very faint and close to the above pre-outburst SDSS brightness.  In addition, ABR08 presented photometry from 2007 February which showed V1647~Ori still close to its pre-outburst brightness (r'=23.26$\pm$0.15).

\subsubsection{The GMOS and SDSS photometric systems}
\begin{figure}[tb] 
\epsscale{0.9}
\plotone{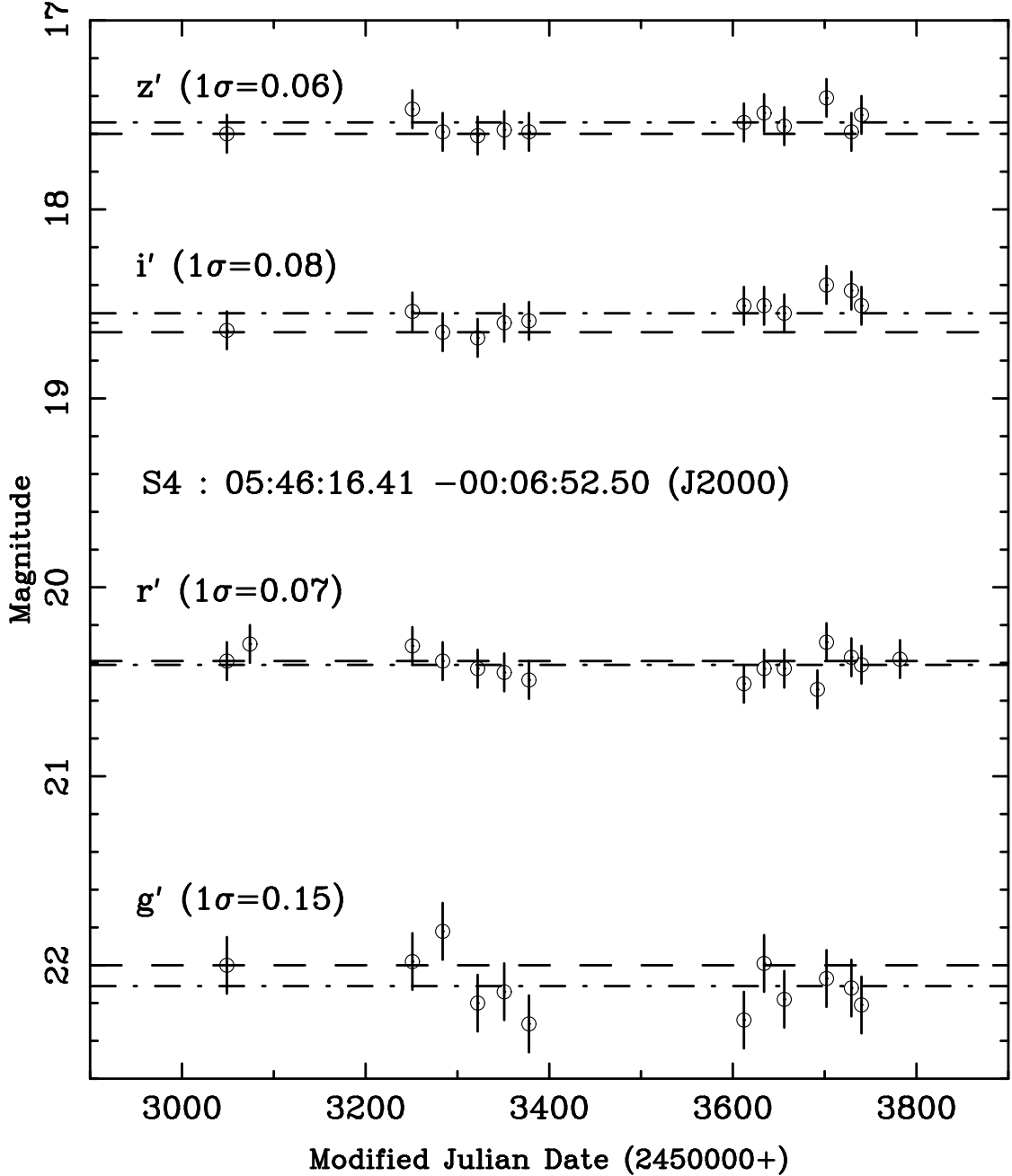} 
\caption{
Optical photometry at g', r', i', and z' of the field star labeled S4 
in Fig.~\ref{idplot}.  The observations are shown as open circles with 
associated error bars.  These values were derived using the SDSS photometry 
of all six calibration stars (see Fig.~\ref{idplot}) as
photometric reference.  For each passband, the associated 1$\sigma$
variation on the mean of all observations is shown assuming that the
star S4 is not intrinsically variable.  The dot-dashed lines show the 
mean value of the aforementioned data points, while the dashed lines 
show the magnitude of S4
derived using the GMOS zeropoints (ZP(g')=27.93, ZP(r')=28.18,
ZP(i')=27.90, ZP(z')=26.77 from Jorgensen 2009) from early 2004.
We note that the difference between the two lines is, in all cases,
smaller than the associated photometric errors. 
\label{sestar}}
\end{figure}

Below, we make use of SDSS photometry and therefore we briefly consider the significance of any differences in photometric systems involved to give confidence to the analysis that follows.  As we noted above, the GMOS g',r',i',z' filters were designed to be close to identical to those used by the SDSS survey (Smith et al. 2002).  We therefore expect the photometric results from GMOS to be quite similar to those from SDSS.  However, to quantify this we have studied the GMOS and SDSS photometry of the calibration sequence used to boot-strap the photometry of V1647~Ori over the monitoring period.  In Fig.~\ref{sestar} we show the photometry of one representative field star labeled S4 in Fig.~\ref{idplot}.  These photometric values were derived using standard GMOS zeropoints for the observing period (i.e. ZP(g')=27.93, ZP(r')=28.18, ZP(i')=27.90, ZP(z')=26.77, Jorgensen 2009) determined using observations of Landolt (1992) standard stars observed close in time to the UT 2004 February 14, V1647~Ori observations.  The magnitudes of the Landolt standards (in the Johnson-Kron-Cousins photometric system) were transformed to the SDSS photometric system using the relationships derived by Smith et al. (2002).  The horizontal lines through the observations are the mean magnitudes of S4 in each GMOS filter (dot-dashed lines) and the SDSS magnitude of S4 from November 1998 SDSS observations (dashed lines).  We note that the slight shift between these lines suggest small differences in stellar magnitudes, however, the shifts are in all cases smaller than the associated 1$\sigma$ error on the data. We conclude, therefore, that the differences between the GMOS and SDSS photometric systems are not significant for the analysis presented below.

\subsubsection{Optical colors} 
\begin{figure}[tb]
\epsscale{1.0} 
\plotone{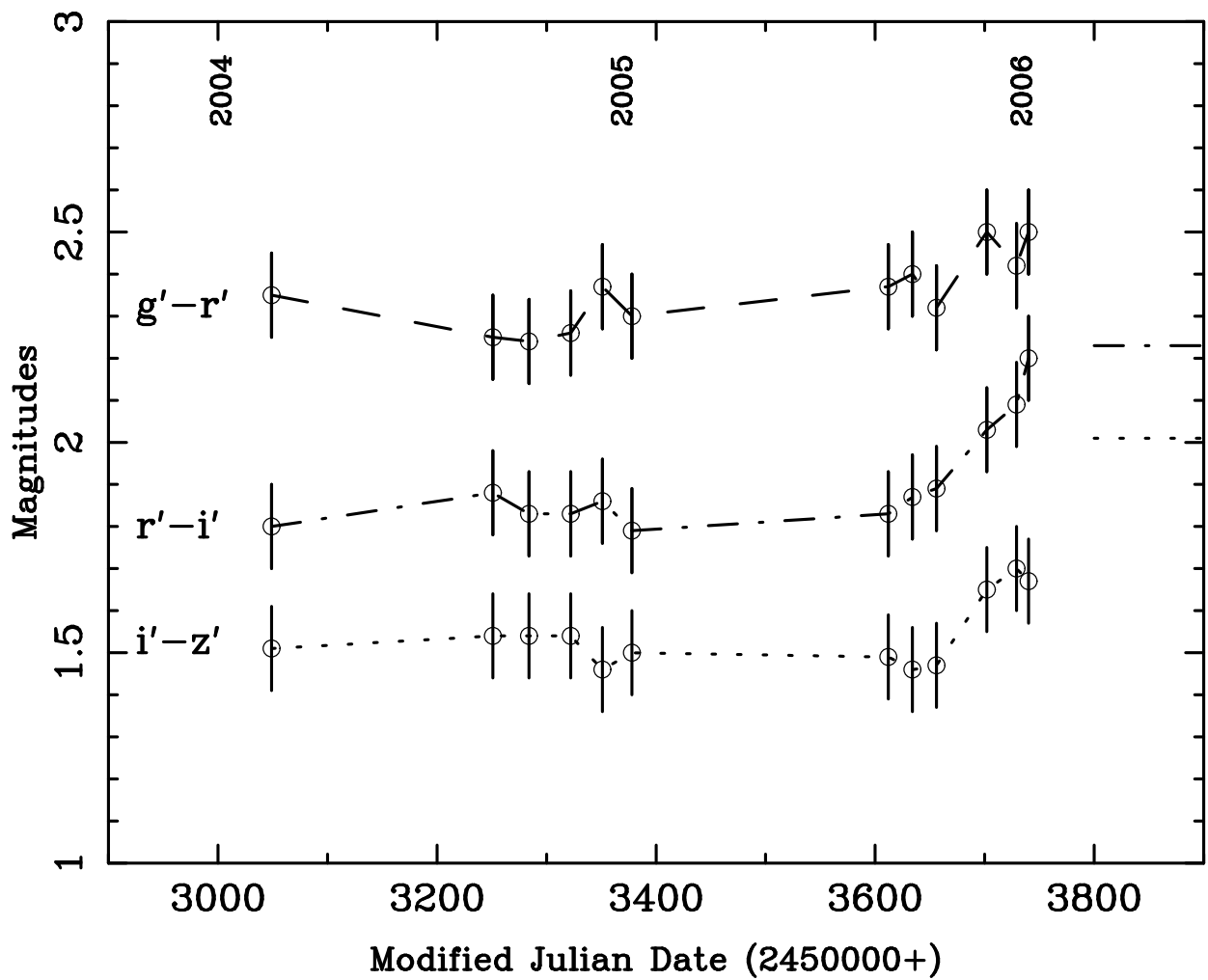} 
\caption{
Optical colors of V1647~Ori spanning the period 2004 February to 2006 
February.  The data were obtained using the Gemini
facility imager/spectrograph GMOS-N in SDSS filters g', r', i', and
z'.  The dashed, dot-dashed, and dotted lines joining the observations
are the g'-r', r'-i', and i'-z' colors, respectively.  The short
dashed, dot-dashed, and dotted horizontal lines from MJD (50000+)
3800--3900 are the pre-outburst colors of V1647~Ori from the SDSS
Orion survey data and additionally presented in McGehee et al. (2004).
\label{colplot}}
\end{figure}

\begin{figure}[tb] 
\epsscale{1.0}
\plotone{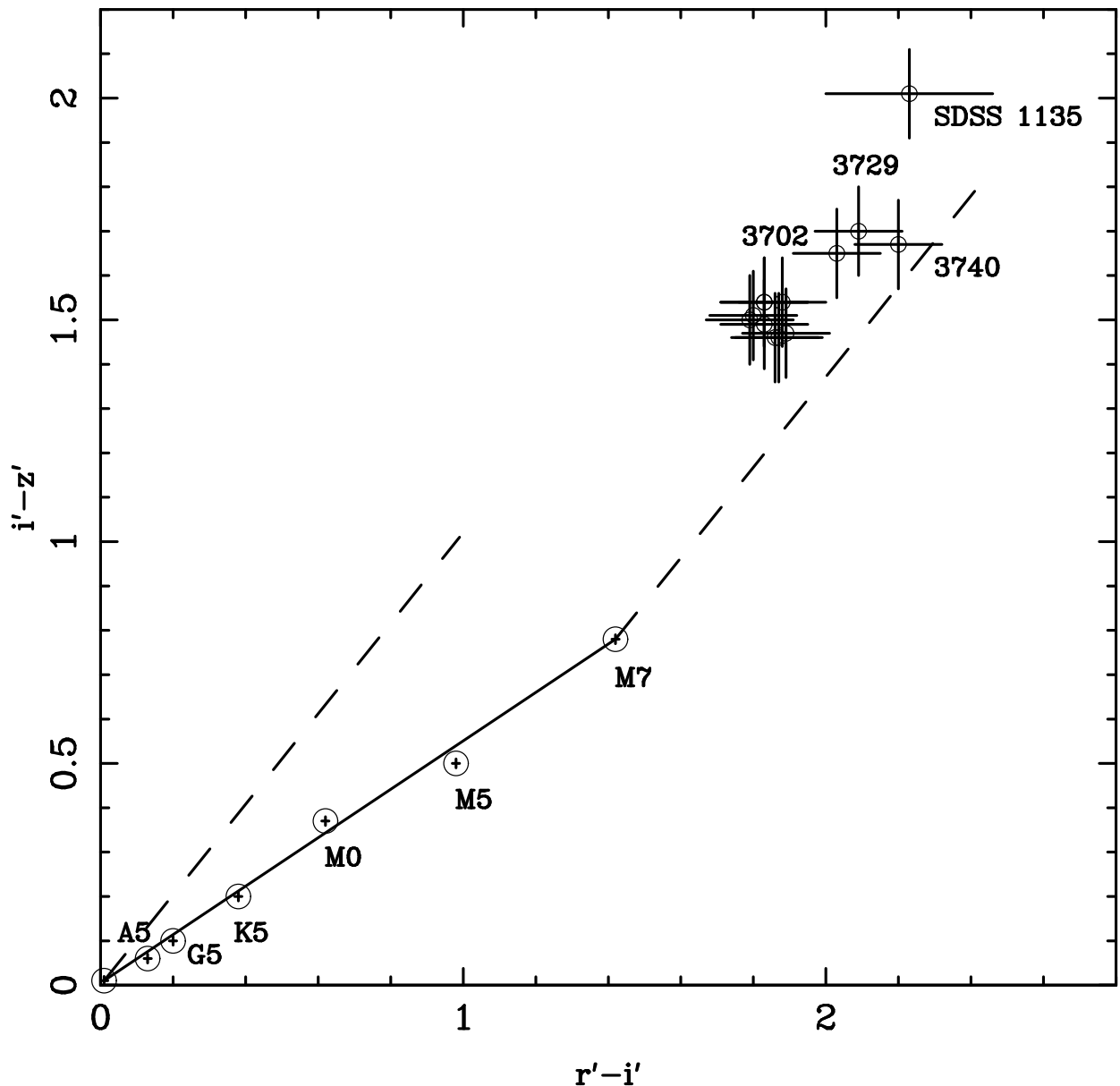} 
\caption{
A color-color plot of the r'-i' and i'-z' colors of V1647~Ori
shown in Fig.~\ref{colplot}.  The solid line is the locus of stellar 
main sequence dwarfs taken from Finlator et al. (2000).  Representative 
spectral types are shown. The dashed lines
extending from the ends of this locus are reddening vectors for R=3.1.
These are calculated using the tabular data provided by D.
Finkbeiner (http://www.astro.princeton.edu/$\sim$dfink/sdssfilters/).  
The three points labeled (MJD 50000+) 3702, 3729,
and 3740 represent data taken close to the end of the outburst.
Their location is consistent with either increased reddening with
respect to the outburst colors, or a change in effective temperature.  
The pre-outburst colors of V1647~Ori are labeled SDSS~1135.
\label{ccplot}}
\end{figure}

\begin{figure}[tb] 
\epsscale{1.0}
\plotone{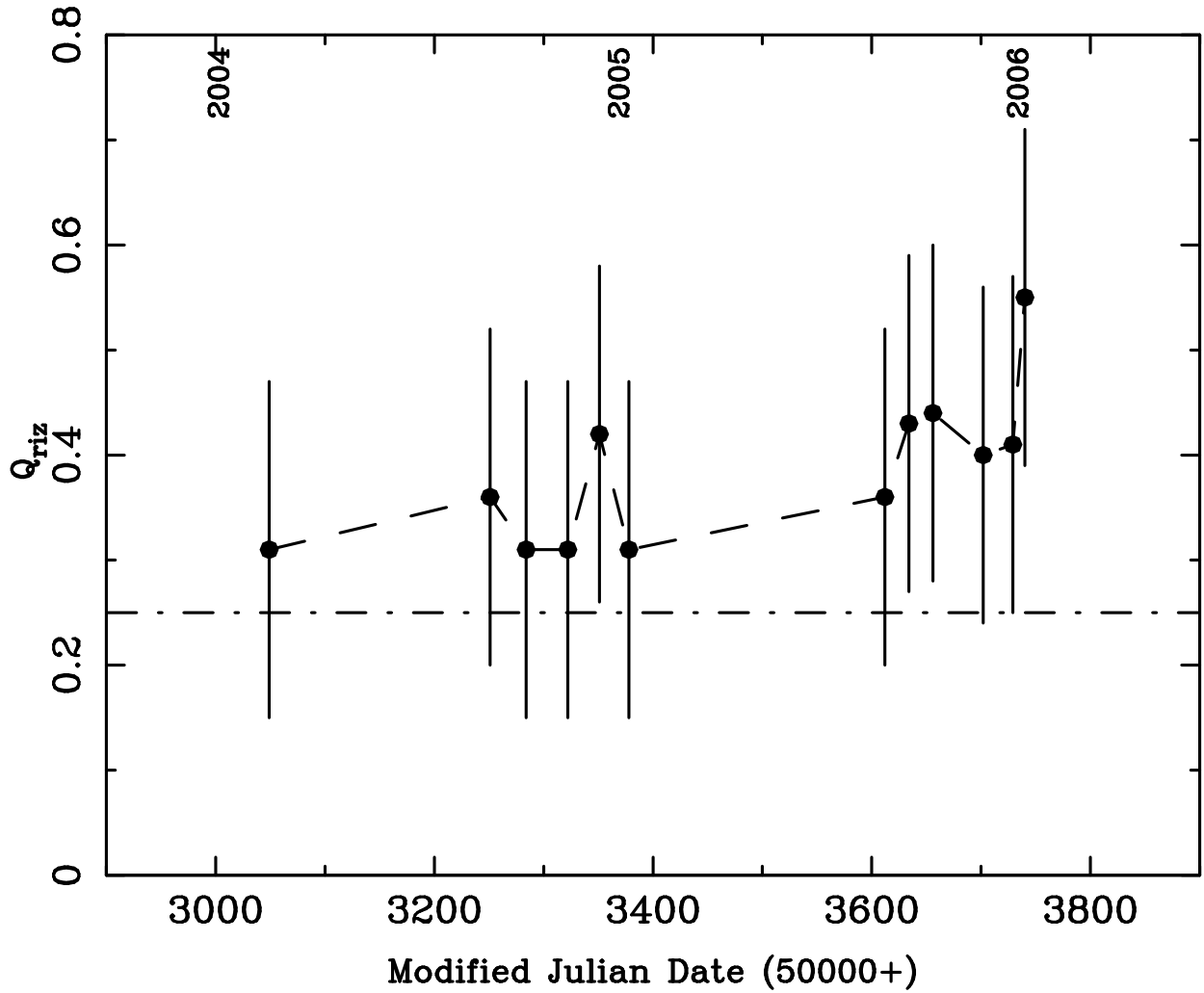} 
\caption{
A plot of the reddening invariant color, Q$_{riz}$, vs. MJD.  Within
the associated errors, the only possible trend in Q$_{riz}$ is a
slight increase in value after MJD~3600.  See Section 3.1.2 for the 
definition of Q$_{riz}$.
\label{colq}}
\end{figure}

During the whole outburst period, when the source was at or close to its maximum brightness, the optical colors remained approximately constant.  In Fig.~\ref{colplot} we show the trend in these colors over the monitoring period.  The available SDSS pre-outburst colors (r'-i' and i'-z') are also shown as horizontal lines at the right edge of the plot.  We note that during the outburst, {\it i)} the optical color indices are larger at shorter wavelengths, {\it ii)} both the r'-i' and i'-z' colors are bluer during outburst (by 0.4 and 0.5 magnitudes, respectively), and {\it iii)} there is a clear trend for these colors returning to their pre-outburst values as the decline phase progresses.  In Fig.~\ref{ccplot} we show the optical r'-i' and i'-z' colors plotted in a two color diagram.  The colors of V1647~Ori after outburst, but before the decline had started, are the cluster of points near r'-i'$\sim$1.8 and i'-z'$\sim$1.5.  The point labeled SDSS~1135 is from Modified Julian Date (MJD 2450000+) 1135 which is UT 1998 November 17 (the pre-outburst SDSS source color).  The three points labeled 3702, 3729, and 3740 are the source colors at those MJDs and are from the decline phase. These points are significantly closer to the pre-outburst color than those during the outburst.  In this plot, we also show the locus of main sequence dwarf colors (solid line) adapted from Fig.~1 of Finlator et al. (2000) and reddening vectors (dashed lines) for a ratio of total to selective absorption R=A$_V$/E(B-V)=3.1 extending from the extremities of the dwarf locus.  These vectors represent a visual extinction A$_V$=5 magnitudes.  We have used the tabular data of D.~Finkbeiner (private communication)\footnotemark \footnotetext{http://www.astro.princeton.edu/$\sim$dfink/sdssfilters/} to calculate the effective change in r'-i' and i'-z' colors over the A$_V$ range plotted.  All the colors of V1647~Ori lie between the two reddening vectors. Dereddening the SDSS~1135 color into the dwarf locus suggests a spectral type of late-K to early M dwarf.  During the outburst phase and through the decline phase, the colors of V1647~Ori first become more blue and then more red.  This can be explained by either one, or a combination, of effects.  Either a change in intrinsic source color or a color {\it and} extinction change has occurred.  We explore this effect in a plot of reddening invariant colors vs. time shown in Fig.~\ref{colq}.  Reddening invariant colors are, as the name suggests, colors that do not change as extinction along the line-of-sight changes.  These have been used by McGehee et al. (2004) in their study of V1647~Ori using SDSS photometry and McGehee et al. (2004) in a study of accretion in low-mass young stars.  Following Table~4 of McGehee et al. (2004), the reddening invariant color we use, Q$_{riz}$ is defined as

\begin{equation}
Q_{riz} = (r'-i')-0.987(i'-z')~~~for~R_V=3.1
\end{equation}

Here we chose the standard interstellar value of R$_V$=3.1, however, we note that the change in Q$_{riz}$ for somewhat larger grains, e.g. R$_V$=5.5, is relatively small ($\sim$10\%) when compared to the associated uncertainties on the measurements.  In Fig.~\ref{colq} we perhaps see a slight trend in Q$_{riz}$ suggesting a larger value as the eruption proceeds into the decline phase. However, the change in value is a small effect with respect to the associated errors.  If this difference is real, it would suggest that the intrinsic colors of the source are different out of outburst than in outburst.  Such a color change is anticipated due to the added optical continuum luminosity from the enhanced accretion during the outburst (McGehee et al. 2004).  Fig.~2 of McGehee et al. (2005) plotted numerical model predictions for the Q$_{riz}$ of low-mass young stars vs. effective temperature, T$_{eff}$, and as a function of surface gravity, log(g).  McGehee (private communication) has subsequently produced a version of this plot for us showing the variation in Q$_{riz}$ up to T$_{eff}$=10000~K.  We do not wish to over-interpret the current dataset due to the associated errors, however, comparing the Q$_{riz}$ value with the aforementioned plot, suggests that in its eruptive state the T$_{eff}$ of the optical emission from V1647~Ori determined from its observed Q$_{riz}$ value ($\sim$0.35$\pm$0.15) is in the range 3000--4000~K.  Distinguishing between different surface gravities in this plot would be impossible since the range of Q$_{riz}$ values for log(g) of 3.5 and 5.5 at this T$_{eff}$ is 0.25 to 0.35, respectively.  Unfortunately, we note that the SDSS pre-eruption value of Q$_{riz}$ from McGehee et al. (2004)\footnotemark \footnotetext{Table~5 from McGehee et
  al. (2004) quotes an eruptive value for Q$_{riz}$ for V1647~Ori of
  0.81.  This used preliminary photometry from Reipurth \& Aspin
  (2004). After a more thorough study of the photometry presented in
  that paper, the true eruptive phase value for Q$_{riz}$ is 0.35.}  is of relatively poor quality (Q$_{riz}$=0.25$\pm$0.23) due to the faintness of V1647~Ori in r' in November 1998 (MJD~1135, r'$\sim$23.04$\pm$0.22) and is therefore not considered further.

\subsubsection{The pre-outburst to post-outburst light curve} 
\begin{figure}[tb] 
\epsscale{0.8}
\plotone{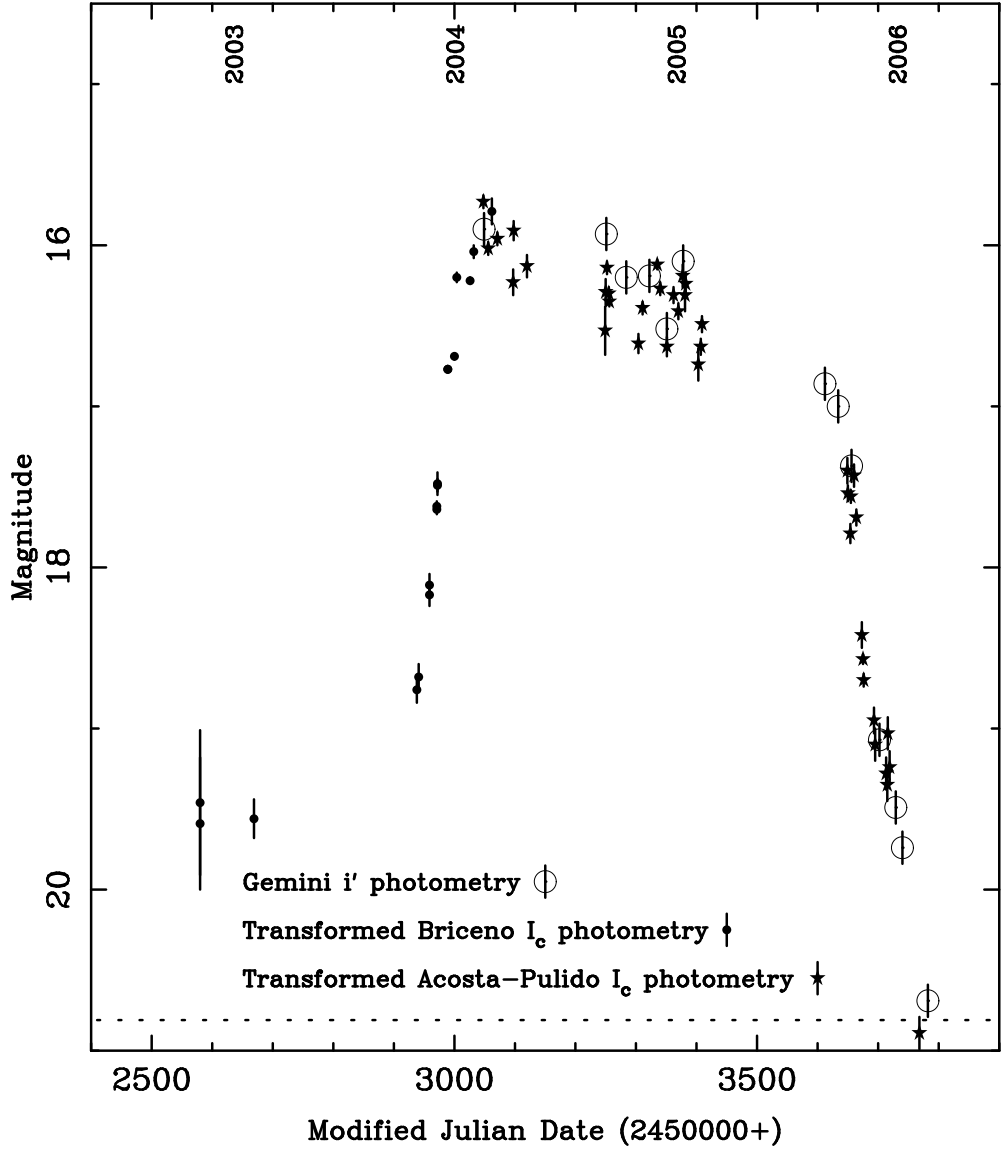} 
\caption{
Optical i' band photometry of V1647~Ori using GMOS (open
circles) from 2004 February to 2006 February together with the I$_c$
photometry from Brice\~no et al. 2004 (filled circles) from 
2003 November to 2004 February, and the Acosta-Pulido et al. (2007) 
I$_c$ photometry (filled stars) from 2004 February to 2006 February.  
The Brice\~no et al. photometry has been transformed from I$_C$ to 
SDSS i' magnitudes using the transformation equations presented in 
Ivezi\'c et al. (2007).  Additionally, the Brice\~no et al. data points 
from MJD 2989 onwards have had an aperture correction applied (+0.62 
magnitudes) since the software aperture used by Brice\~no et al. had 
a radius of 4$\farcs$1 while our aperture radius was 1$''$.  The 
horizontal black dotted line shows the November 1998 SDSS i' photometric 
magnitude of the source (i'=20.81). Prior to the time period displayed, 
Brice\~no et al. reported the I$_c$ magnitude of V1647~Ori to be 
18.44$\pm$0.11 (i'=19.31) in January 1999, and 20.08$\pm$0.3 
(i'=20.95) in December 1999.
\label{icphotplot}}
\end{figure}

In order to produce as complete a light curve of the 2003--2006 outburst of V1647~Ori as possible, we have combined our photometry with selected published results.  In Fig.~\ref{icphotplot} we show our SDSS i' band GMOS photometry from Table~\ref{photom} together with the I$_C$ photometry of Brice\~no et al. (2004).  We have used the polynomial transformation equation
 of Ivezi\'c et al. (2007) to estimate SDSS photometric i' values from the Kron-Cousins (Landolt 1983) I$_C$ photometry of Brice\~no et al. (2004).  The transformation equations used was

\begin{equation}
i' = I_C + 0.0307(r'-i')^3 - 0.1163(r'-i')^2 + 0.3341(r'-i') + 0.3584
\end{equation}

Since this (and any) transformation is source color dependent, we assume an intrinsic color for V1647~Ori of r'-i'=1.9, the mean of the pre-outburst and outburst colors.  This results in a transformation correction of 0.79 magnitudes.  Additionally, since the software aperture used by Brice\~no et al. is considerably larger than ours (aperture radii of 4$\farcs$1 and 1$\farcs$0, respectively), we have estimated from our UT 2004 February 14 image an aperture correction for the transformed Brice\~no et al. data.  Since the nebulosity local to V1647~Ori was reasonably bright in Brice\~no et al. images from 2004 December 15 onwards, we apply the aperture correction to only their data points subsequent to this date. The correction used was 0.62 magnitudes. The result of the transformation and aperture correction is to give a reasonably good match ($\Delta$m=0.1$^{m}$) between our photometry (from MJD~3049) and the closest value from Brice\~no et al. (from MJD~3036).  Also shown in this figure is the I$_C$ photometry from Acosta-Pulido et al. (2007).  We have used the same transformation to SDSS photometry given by Ivezi\'c et al. (2007) and shown in Eqn.~2 above.  In Fig.~\ref{icphotplot} our data are shown as open circles, the Brice\~no et al. data as filled dots, and the Acosta-Pulido et al. data as filled stars.  We additionally note that the pre-outburst (MJD~1135) SDSS i' photometry from McGehee et al. (2004) (indicated by the black dotted line in Fig.~\ref{icphotplot}) is consistent with our last data point (MJD~3782) suggesting that V1647~Ori had more or less returned to its pre-outburst brightness in 2006 February.

\begin{figure}[tb] 
\epsscale{0.8}
\plotone{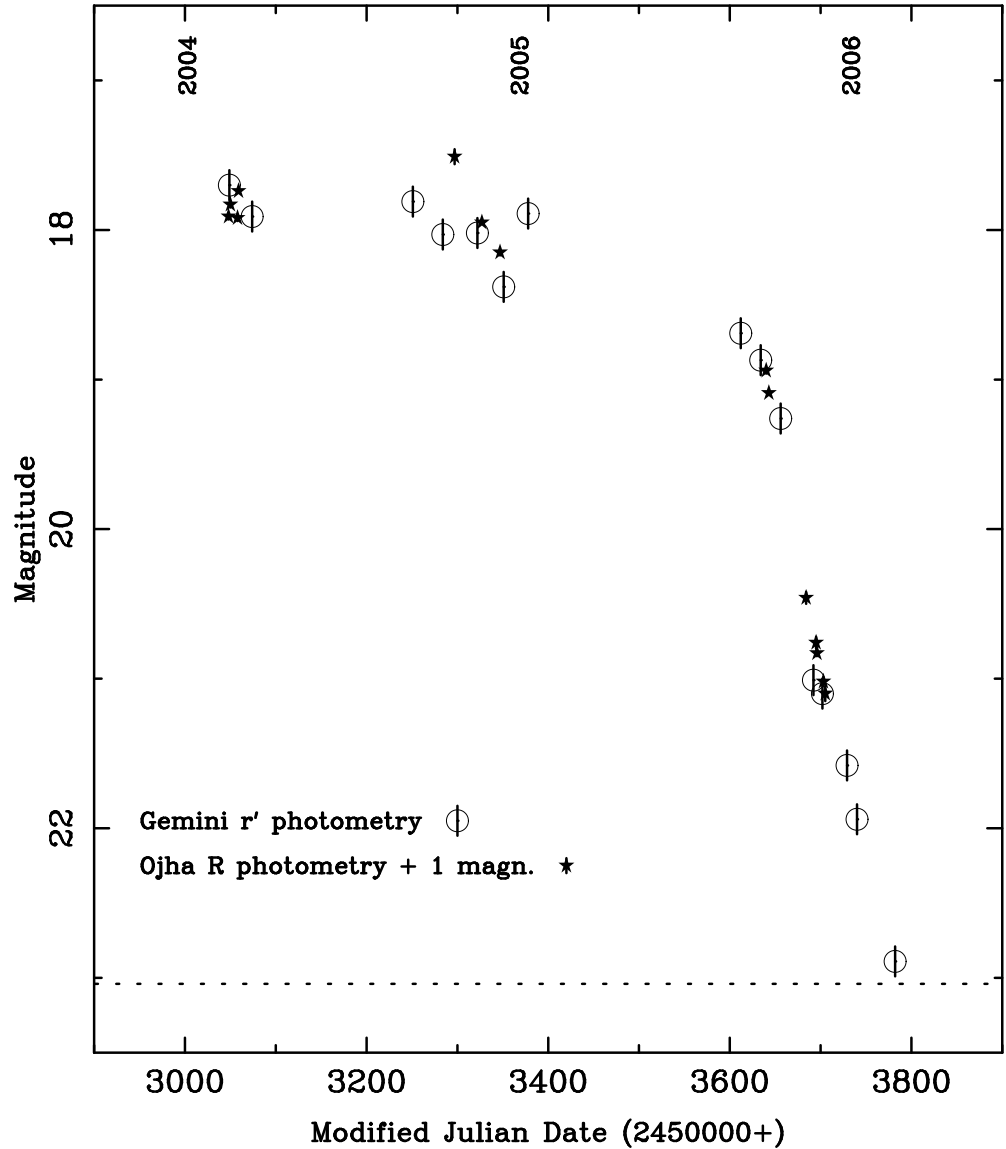} 
\caption{
Optical r' band photometry of V1647~Ori using GMOS (open circles)
from 2004 February to 2006 February together with the R photometry
from Ojha et al. 2006 (filled stars) from 2004 February to 2005 November.   
We have shifted the Ojha et al. data vertically by 1 magnitude
to obtain good correspondence between the Gemini and Ojha et
al. values throughout the overlap region.  
The horizontal dotted line is the November 1998 SDSS r' photometry 
(r'=23.04).
\label{rphotplot}}
\end{figure}

Another dataset that has good temporal coverage using only one telescope/instrument combination is the R band photometry of Ojha et al. (2006).  They present 13 data points covering the period from peak brightness to approximately half way down the steep decline phase.  These data are shown in relation to our r' band photometry in Fig.~\ref{rphotplot}.  Due to the extensive overlap of data, we have made no attempt to match the two photometric systems and have merely shifted the Ojha et al. data (filled dots) by $+$1 magnitude (fainter) to match our data (open circles) as well as possible.  We again note that our final r' photometric point (MJD~3782) is very close in value to the pre-outburst (MJD~1135) SDSS value of 23.04$\pm$0.22 from McGehee et al. (2004).

\begin{figure}[tb] 
\epsscale{0.8}
\plotone{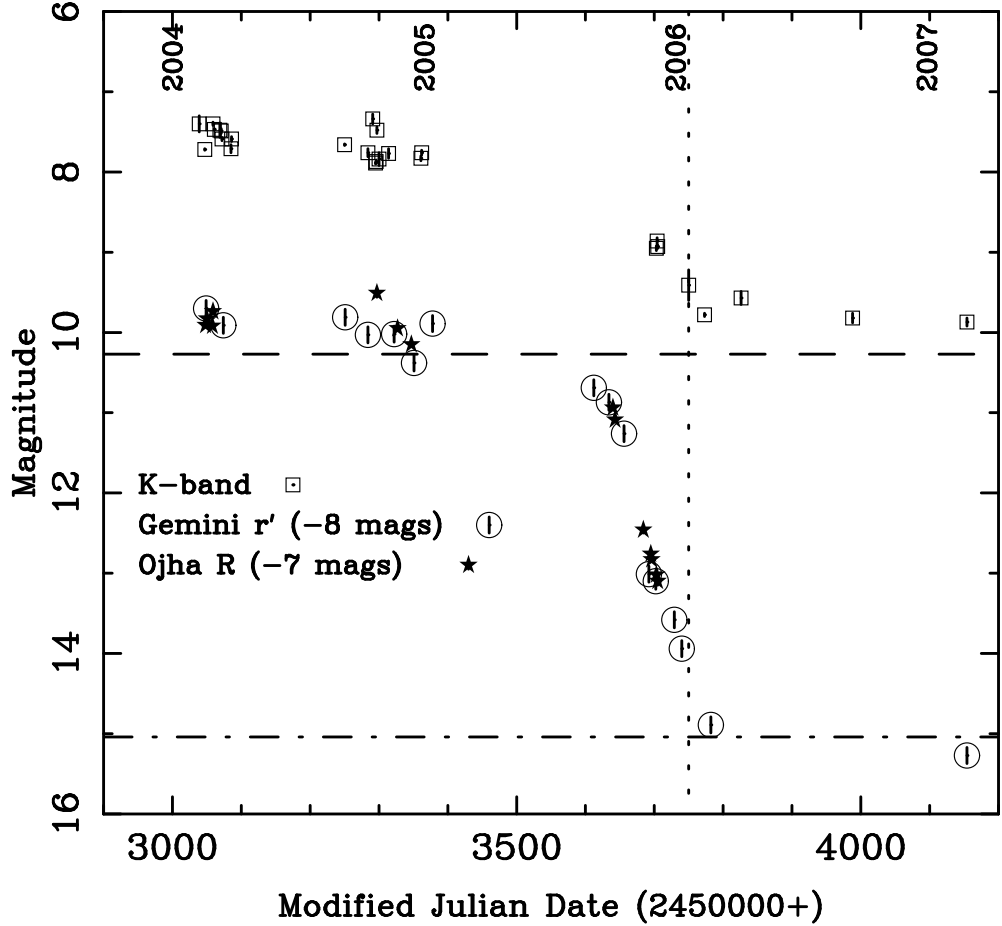} 
\caption{
The optical r' band light-curve from Fig.~\ref{rphotplot} together 
with the near-IR K-band light-curve compiled from the literature.  
The r' band photometry is shown as open circles (this paper) and 
filled stars (Ojha et al. 2006).  The K-band photometry is shown 
as open squares (from Acosta-Pulido et al. (2007) and Ojha et al. 
2006).  We have shifted the optical data vertically by 8 magnitude 
so that both light-curves can easily been  compared.  The horizontal 
dot-dashed line is the November 1998 SDSS r' photometry (r'=23.04).  
The horizontal dashed line is the October 1998 2MASS K-band photometry 
(K=10.27).  The vertical dotted line is at MJD~3750 and is merely 
there as an aid in relating the two light-curves.
\label{rkplot}}
\end{figure}

We have additionally compiled all published near-IR K-band (2~$\mu$m) photometry and plotted it together with the above r' light curve in Fig.~\ref{rkplot}.  This comparison shows that in the near-IR the change in brightness of V1647~Ori was less than in the optical and amounted to $\sim$2.8 magnitudes from the October 1998 2MASS photometry (horizontal dashed line).  The data were taken from Acosta-Pulido et al. (2007) and Ojha et al. (2006) although some of the data points in Ojha et al. (2006) were taken from other sources (specifically Reipurth \& Aspin 2004, McGehee et al. 2004, and Ojha et al. 2005).  Unfortunately, there is no K-band photometry from the spring and early fall of 2006 and therefore we cannot, with certainty, say whether the two light-curves are in phase or if there is a relative delay between them.  The few points in late 2006 and early 2007 suggest that the major decline phase of V1647~Ori was somewhat shallower at K than in the optical although this relies on so few points that it is by no means certain.  Nonetheless, the main fact arising from this plot is that the amplitude of the K-band outburst was considerably less (by 2.2 magnitudes) than in the optical.
%This implies that the luminosity from the accretion had more of an effect at optical wavelengths than in the near-IR and that perhaps the increase in K-band flux is from an increased K-band thermal excess (dust heated by the accretion flux) rather than the accretion emission itself.

\begin{figure}[tb] 
\epsscale{0.8}
\plotone{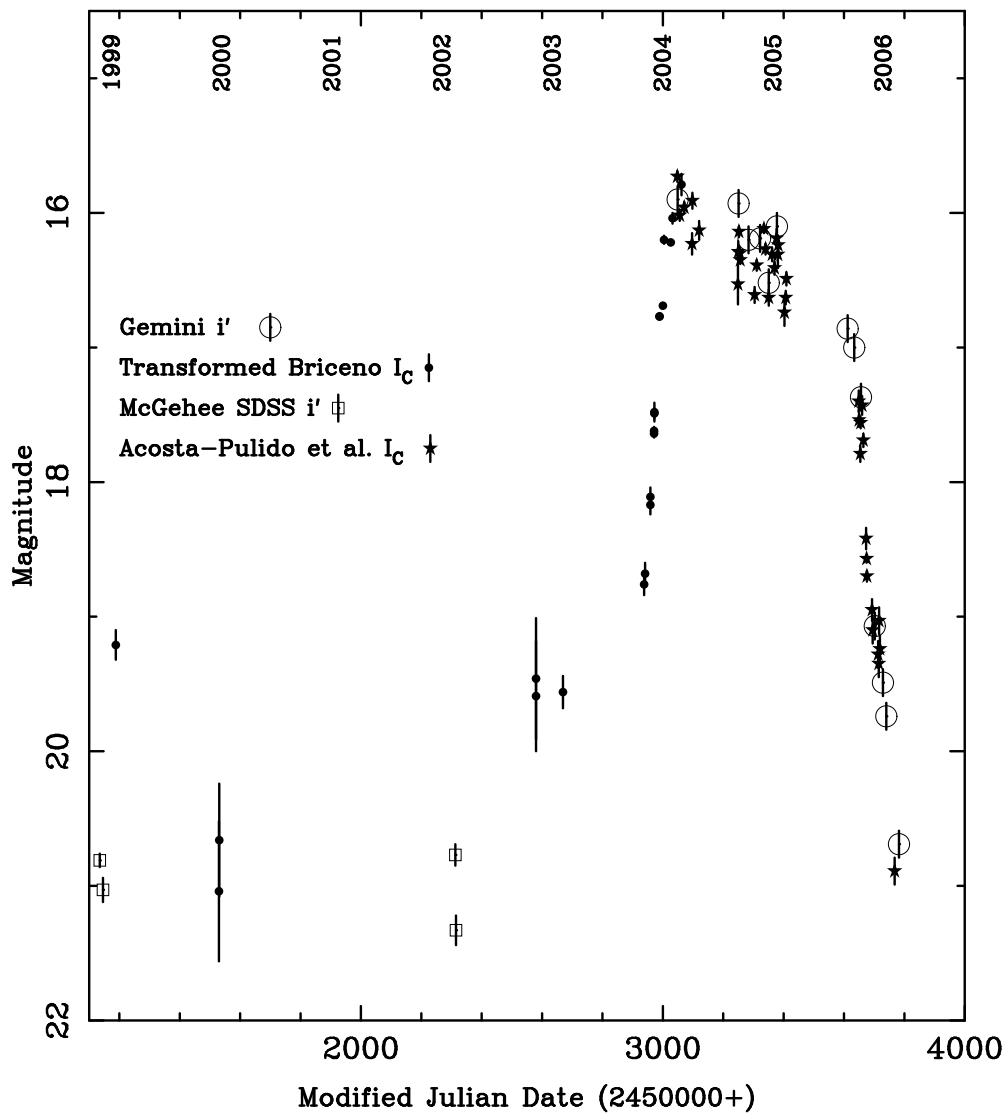} 
\caption{
Optical photometry of V1647~Ori using a) GMOS i' band data (open 
circles) from 2004 February to 2006 February, b) the Brice\~no et al. 
(2004) I$_c$ band data transformed to i' magnitudes (filled dots) 
from 1999 January to 2004 February, c) the i' band SDSS data from 
McGehee 
et al. (2004) (open squares) from 1998 November to 2002 February, 
and d) the Acosta-Pulido et al. (2007) I$_c$ band data transformed 
to i' magnitudes (filled stars) from 2004 February to 2006 February.  
The last point of the GMOS data has been estimated from the r' band 
observation on that MJD using the r'-i' colors of the previous 
observation (r'-i'=2.2).
\label{ifullplot}}
\end{figure}

From Figs.~\ref{icphotplot}, \ref{rphotplot} and \ref{rkplot}, and previously published data, we can conclude that:

\begin{itemize}
\item The total duration of the eruptive event was between 847 and 932
  days ($\sim$28 and 31 months or 2.3 to 2.6 years).  This value
  assumes the outburst began between MJD~2850 and 2935 and ended
  around MJD~3782.

\item The pre-outburst (SDSS, MJD~1135) and post-outburst (Gemini,
  MJD~3782) brightness of V1647~Ori are very similar.

\item The rise from the pre-outburst brightness to peak brightness
  took between 145 and 230 days (due to the uncertainty in the date of
  the start of the eruption, see above).

\item The decline from peak brightness to pre-outburst brightness had
  two well-defined regions, a shallower phase from the late summer of
  2004 to the fall of 2005 ($\sim$410 days) when V1647~Ori faded by
  $\sim$1 magnitude, and a steeper phase from the fall of 2005 to its
  pre-outburst brightness in early 2006 ($\sim$130 days) when it faded
  by an additional $\sim$3.5 magnitudes.

\item Short-term variability was present during the shallow decline
  phase.  This variability amounted to a peak-to-peak amplitude of
  $\sim$0.7 magnitudes.

\item The 2003 eruptive event may have started from a somewhat
  elevated state.  In November 1998 (MJD~1135) the SDSS (r' and i')
  photometry lies between 1 and 2 magnitude fainter than the
  brightness when the major eruption phase commenced in late 2003.  In
  addition, our last r' observation from February 2006 indicates that
  the star had returned to this fainter level after the outburst had
  ceased.  This behavior is shown graphically in Fig.~\ref{ifullplot}
  where all of the above optical i' and I$_C$ data are plotted
  together with an interpolated i' magnitude from 2006
  February.\footnotemark \footnotetext{using the 2006 February r'
    magnitude and the r'-i' color from our 2006 January observation.}
  We note also that the two I$_C$ observations of Brice\~no et
  al. (2004) from 1999 suggest that the source possessed significant
  variability when faint with photometry from January 1999
  (I$_C$=18.44$\pm$0.11 transforms to i'=19.31) and December 1999
  (I$_C$=20.08$\pm$0.4 transforms to i'=20.95) exhibiting a
  statistically significant difference of $\sim$1.6 magnitudes.  This
  may suggest that V1647~Ori underwent a period of significant
  instability prior to the main eruption phase.

\item The optical (r') and near-IR (K) light-curves differ
  significantly in their amplitude with the optical being around 2.2
  magnitudes larger.
%This is consistent with the enhanced accretion event producing mostly UV/optical radiation as indicated by the B spectral type inferred from the reflected light spectrum by Brice\~no et al. (2004).  Perhaps the K-band excess during outburst was mostly created by dust in the circumstellar environment (disk) heated by the hot accretion flux. 

\item Some 12 months after the decline to pre-outburst brightness,
  V1647~Ori remained optically very faint and close to its
  pre-outburst brightness.
\end{itemize}

\subsection{Optical Spectroscopic Features and their Evolution}
In this section, we discuss our optical spectroscopic observations over the monitoring period but defer discussion of the strongest feature in the spectra, the H$\alpha$ line, to $\S $3.3.  However, we do consider the H$\alpha$ line when describing the HIRES echelle spectra below as an introduction to the line properties.  We note that all radial velocities quoted below have been corrected to be heliocentric and that to convert to velocities in the stars frame of reference the reader should use the rest velocity of the L1630 molecular cloud measured directly on V1647~Ori by Andrews, Rothberg, \& Simon (2004) to be +10~km~s$^{-1}$ (see their Fig.~3). 

\begin{figure}[tb] 
\begin{center}
\includegraphics*[angle=0,scale=1.0]{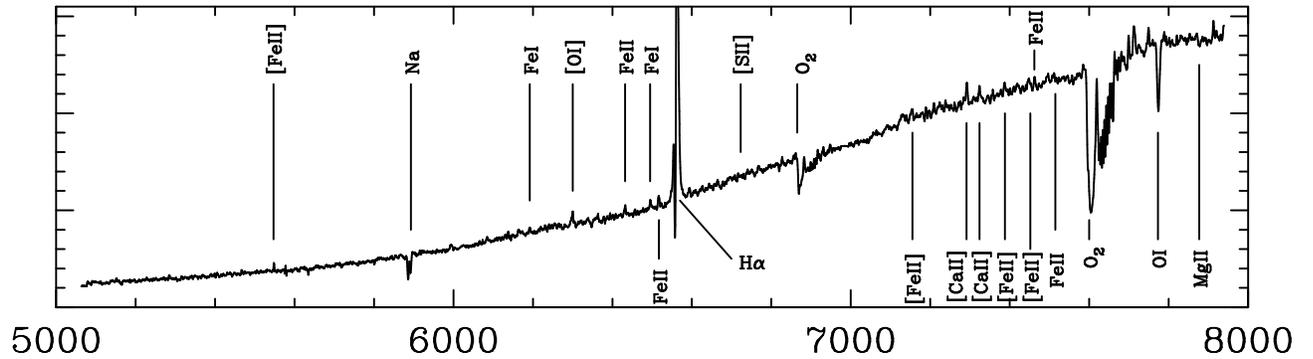} 
%\epsscale{1.0}
%\plotone{2004oct06-sm.ps} 
\caption{
Optical GMOS spectrum of V1647~Ori from 2004 October 6.  This is 
a representative spectrum of the 15 acquired between February 2004 
and February 2006.  In 2004 October, V1647~Ori had a magnitude of 
r'=18.03 and was in its shallow decline phase.  All identifiable 
absorption and emission lines are marked including the telluric 
O$_{2}$ and O2-A bands.  Note the prevalence of weak permitted and
forbidden Fe emission lines.  The major intrinsic feature is the 
strong H$\alpha$ emission line with a blue-shifted absorption 
component forming a P~Cygni type profile.  Table~\ref{lineids} 
shows the variation in spectral features over the monitoring period.
\label{oct04plot}}
\end{center}
\end{figure}

\begin{figure}[tb] 
\epsscale{1.0}
\plotone{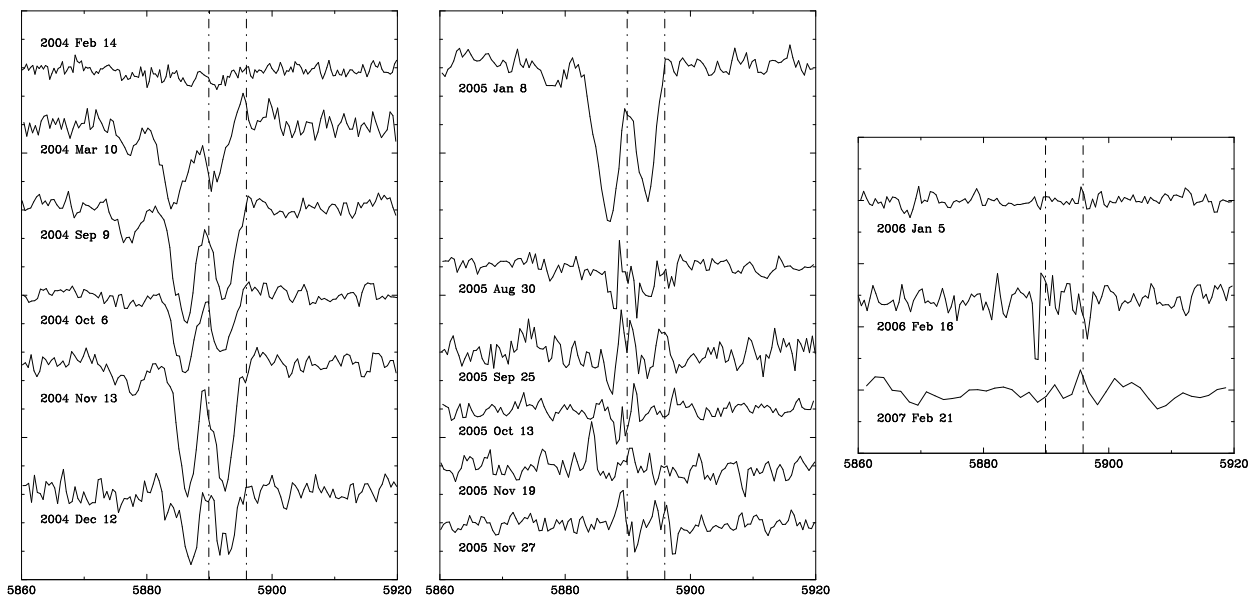} 
\caption{
Variation in Na~D 5890 and 5896~\AA\ absorption profiles over the 
monitoring period.  The continuum has been subtracted in these 
spectra and they have not been smoothed so as to preserve the 
intrinsic noise structure for comparative purposes.  The vertical 
dashed lines show the Na lines rest wavelengths.  In 2004 February, 
the Na lines were barely visible but soon became very pronounced 
and blue-shifted with respect to the rest wavelength.  The lines 
are highly variable in depth and shape until 2005 August (at the 
beginning of the steep decline phase) when they had faded 
considerably.  For the remainder of the monitoring period (2005 
September to 2007 February) the lines were barely present.
\label{nad-all}}
\end{figure}

\begin{figure}[tb] 
\epsscale{1.0}
\plotone{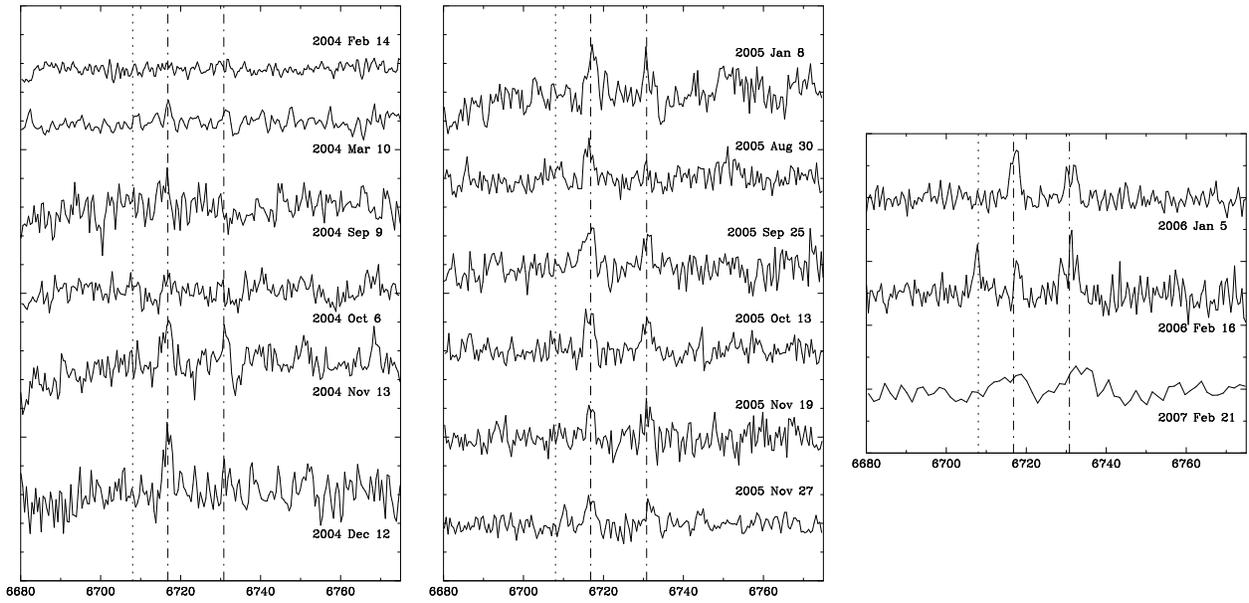} 
\caption{
Variation in the shock-excited [S~II] 6717 and 6731~\AA\ emission 
line intensity and ratios over the monitoring period.  The vertical 
dashed lines show the [S~II] lines rest wavelengths.  In 2004 
February, the [S~II] lines were not present.  In the spectrum 
taken on 2004 March 10 however, they are seen faintly.  The 
intensity and ratio of the lines varies from then to the end of 
the monitoring period.  Again, the spectra have been continuum 
subtracted and not smoothed.  
\label{sii-all}}
\end{figure}

We show a GMOS spectrum of V1647~Ori in Fig.~\ref{oct04plot} selected as representative of the 15 GMOS spectra we obtained between February 2004 and February 2006.  These are the data from 2004 October 6 since it has good signal to noise and shows most of the spectral features present.  The source shows a rising red continuum with strong H$\alpha$ emission exhibiting a P~Cygni profile.  There are numerous weak features in the spectrum including the Na~D absorption at 5890 and 5896~\AA.  Table~\ref{lineids} lists the lines found in the spectra and their variability over the monitoring period.  We have not measured equivalent widths for the lines since, as we shall discuss below, such a value is affected significantly by the highly variable optical continuum flux observed.  Instead, we have merely specified whether the lines are in absorption or emission, or not present.  As well as permitted emission lines of Fe, O, and Mg, numerous forbidden emission features, i.e. [Fe~II], [O~I], [S~II], and [Ca~II], are observed.  In Figs.~\ref{nad-all}, \ref{sii-all}, \ref{6300OI-all}, \ref{fe-all}, and \ref{caii-all}, we show extracted wavelength ranges around the Na~D absorption lines, and the [S~II], [O~I] (6300~\AA), Fe (6400--6550~\AA), and the Ca emission lines (7200--7400~\AA), respectively.

\subsubsection{Na~D absorption lines}
The Na~D neutral absorption line profiles are shown in Fig.~\ref{nad-all}.  The first observation of the monitoring period, taken on 2004 February 14, only showed weak Na~D absorption.  This was blue-shifted by --200~km~s$^{-1}$ with respect to the rest wavelengths of the lines.  Between 2004 February 14 and March 10, the Na line profiles change significantly, they became stronger and considerably broader.  On 2004 March 10, they were blue-shifted by --280~km~s$^{-1}$ with respect to the line rest wavelengths.  Also at this time two emission features, presumably Na~D lines, are present at 5895.3~\AA\ (+270~km~s$^{-1}$ with respect to the bluer Na~D line rest wavelength of 5889.9~\AA) and 5899.7~\AA\ (+190~km~s$^{-1}$ with respect to the redder Na~D line rest wavelength of 5895.9~\AA).  These appear to be real features since emission is also seen in the lower resolution 2004 February 18 spectrum from Brice\~no et al. (2004)\footnotemark\footnotetext{Kindly made available to us by
  C.~Brice\~no.}.  An absorption feature is seen at 5877~\AA\ and its depth appears correlated to the strength of the Na~D lines.  This absorption feature is similar in width ($\sim$160~km~s$^{-1}$) to the main Na~I lines and could possibly be a highly blue-shifted component of Na absorption appearing at --640~km~s$^{-1}$ with respect to the shorter wavelength Na~I line at 5890~\AA.  An alternative explanation for the presence of this line is that it is the He~I line at 5876~\AA\ in absorption.  Perhaps this is the more reasonable interpretation and is supported by the fact that we also observe He~I lines at 6678 and 7766~\AA\ in absorption.

After the summer months of 2004, the first spectrum taken was on 2004 September 9 and showed the Na~D lines even stronger than on March 10.  Between September 9 and 2005 January 8 (the last observation before the 2005 summer break), the Na lines vary in intensity, becoming weaker and stronger in successive months.  The Na lines appear strongest on 2005 January 8 and are blue-shifted with respect to the rest wavelength by $\sim$--150~km~s$^{-1}$.  Large changes occur between 2005 January 8 and the next observation on 2005 August 30 where the Na lines have weakened significantly and appear less blue-shifted ($\sim$--100~km~s$^{-1}$).  In the UT 2005 September 25 spectrum, the Na lines possibly appear weakly in emission and very close to their rest wavelengths (--40~km~s$^{-1}$).  From 2005 October 13 to the end of the monitoring period, the Na lines have more or less disappeared.

Highly structured Na~I 5890 and 5895~\AA\ lines are not uncommon amongst eruptive variables, specifically FUors.  Examples include FU~Ori itself (Bastian \& Mundt 1985; Hartmann \& Calvet 1995), BBW~76 (Reipurth et al. 2003), V1057~Cyg (Bastian \& Mundt 1985; Herbig, Petrov, \& Duemmler 2003), and V1515~Cyg (Bastian \& Mundt 1985).  It was noted by Bastian \& Mundt (1985) that all Na~D profiles appear very similar to each other.  This also is the case if we compare these three FUors to the profile of BBW~76 from Reipurth et al. (2003).  The time-series of BBW~76 by Reipurth et al. (2003) is particularly interesting since they observed multiple minima in both Na lines which were predominantly blue-shifted by up to --300~km~s$^{-1}$.  In addition, between 1985 and 1994 the lines varied significantly in both width and blue-shifted velocity.  We note here that the generally accepted interpretation of such Na~I line structure is that an intense stellar wind, resulting from enhanced accretion, forms shell-like expanding/outflow structures which are accelerated close to the star/disk and slow with increasing distance (Bastian \& Mundt 1985; Reipurth et al. 2003).  Since our data do not have the spectral resolution of the BBW~76 data (R$\sim$2,000 as opposed to 20,000 for BBW~76) we cannot resolve individual shell components. However, the large line width and the blue-shifted nature of the absorption implies in itself that we are seeing considerable outflowing material with a significant velocity gradient.  If the absorption feature at 5877~\AA\ is indeed a high-velocity blue-shifted Na component, then within one month (2004 February 14 to March 10) the wind dramatically increases in density (line depth) and velocity (line width/range) with a maximum of over --600~km~s$^{-1}$.  For a spherically symmetric wind, the fact that we do not observe a symmetric profile, one with both blue-shifted and red-shifted absorption, suggests that we only have an unobscured line-of-sight towards the absorbing material that is expanding towards us.  This is likely the result of obscuration by the circumstellar/accretion disk and implies that the material lies close to the star and is accelerated, in this location, to high velocities (see the discussion of Bastian \& Mundt 1985 for more details).  This is consistent with the inferred inclination of the axis of the nebula (McNeil's Nebula) to the line-of-sight of $\sim$60$^{\circ}$ (Acosta-Pulido et al. 2007; ABR08), since we are looking onto the star/disk through the nebula.

\subsubsection{[S~II] emission lines}
Fedele et al. (2007) and ABR08 found [S~II] emission at 6717 and 6731~\AA\ in the spectrum of V1647~Ori in 2006 January and 2007 February, respectively.  In Fig.~\ref{sii-all}, we trace the evolution of this emission from 2004 February to 2007 February.  We see that there are weak [S~II] lines present as early as 2004 March 10 and they are seen in all our spectra up until the end of the monitoring period.  The lines fall at their rest wavelength with the resolution we have, and vary in both intensity and ratio with time.  The ratio 6717/6731 is greater than unity from 2004 March 10 to 2006 January 5, and then, in our last two spectra, drops to less than unity.  From 2005 October 13 onwards, the lines appear to be double-peaked with a separation of the peaks being $\sim$40~km~s$^{-1}$.  However, our spectral resolution is insufficient to really study this structure in any detail or even to be certain of its existence.  If it is present then perhaps we are detecting red- and blue-shifted shocks from the star.

We have not derived electron densities, {\it n$_{e}$}, from the ratio of the two sulfur lines (6717/6731, the ratio is only weakly dependent on temperature) from all our spectra since the signal to noise of the line detection is, in many cases, insufficient.  However, for five observations, 2005 October 13, 2005 November 27, 2005 December 25, 2006 January 05, and 2007 February 21, we consider the lines well enough detected.  The derived ratios for 6717/6731 are 1.3$\pm$0.1, 1.4$\pm$0.1, 1.4$\pm$0.1, 1.4$\pm$0.1, and 0.8$\pm$0.1 which give values of {\it n$_{e}$}, derived using the IRAF {\tt nebular.temden} program for a temperature of 10$^{4}$~K, of 26--240, $<$121, $<$121, $<$121, and 910--2230~cm$^{-3}$, respectively.  The ranges come from the uncertainties quoted on the line ratios and the upper limit comes from a failure if {\tt temden} to calculate densities for a ratio $>$1.4.  It seems, therefore, that the forbidden sulfur lines are formed in a rarefied gas which becomes denser towards the end of our monitoring period although, even then, it is far from the [S~II] critical density of 20,000~cm$^{-3}$.

\subsubsection{[O~I] 6300\AA\  line}
\begin{figure}[tb] 
\epsscale{1.0}
\plotone{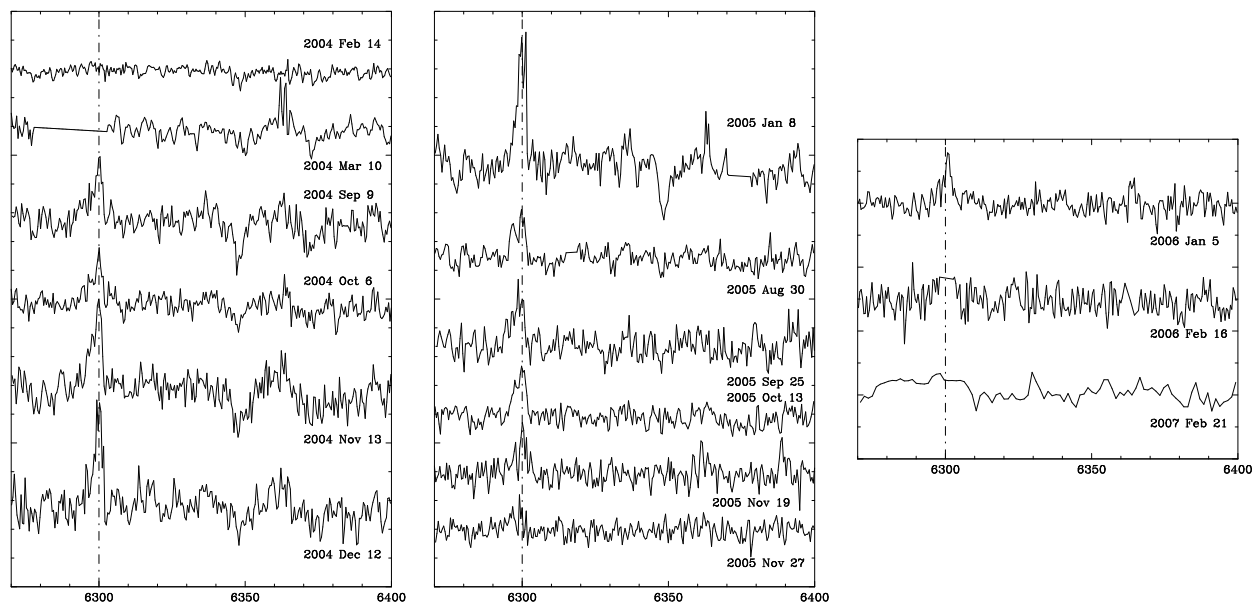} 
\caption{
Variation in the [O~I] 6300\AA\ emission line intensity over the 
monitoring period.  The vertical dashed line shows the [O~I] line 
rest wavelength.  The line is not present on 2004 February 14 but
is strong in the spectrum taken on UT 2004 September 9.  The 
spectrum from 2004 March 10 had considerable noise spikes at 
the wavelength of [O~I] which have been removed by interpolation.  
The line intensity mostly increases from UT 2004 September 9 to 
UT 2005 January 8 when it begins to decline.  The spectrum from 
2006 January 5 had some noise spikes (which have been removed) 
but the wings of the [O~I] line can still be seen.  The line is
perhaps just detected in the UT 2007 February spectrum.  Again, 
the spectra have been continuum subtracted and not smoothed.  
The [O~I] emission line is asymmetric in shape with an extended 
blueshifted wing.  We note the presence of an unidentified 
absorption feature at $\sim$6350~\AA.
\label{6300OI-all}}
\end{figure}

\begin{figure}[tb] 
\epsscale{1.0}
\plotone{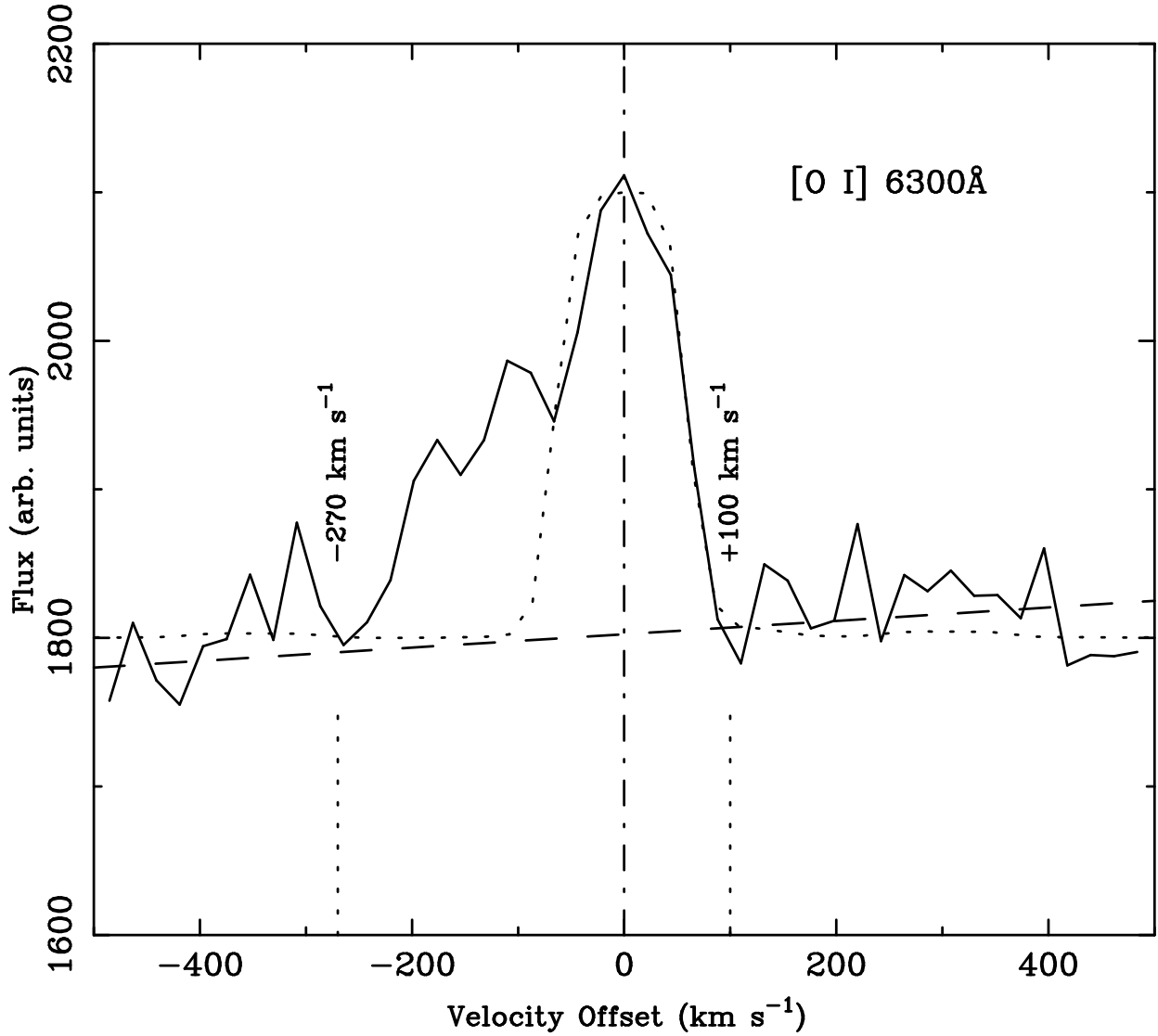} 
\caption{
The [O~I] 6300\AA\ emission line profile from the UT 2004 November 
13 spectrum.  The vertical dot-dashed line shows the [O~I] line 
rest wavelength.  The vertical dotted lines from the bottom axis show the extent of 
the emission profile which corresponds to --270 to $+$100~km~s$^{-1}$.  
The sloping dashed line is the continuum level across the emission feature.
The dotted line under [O~I] shows an arc spectrum lamp line close in wavelength 
to [O~I].
\label{6300OI-v}}
\end{figure}

\begin{figure}[tb] 
\epsscale{1.0}
\plotone{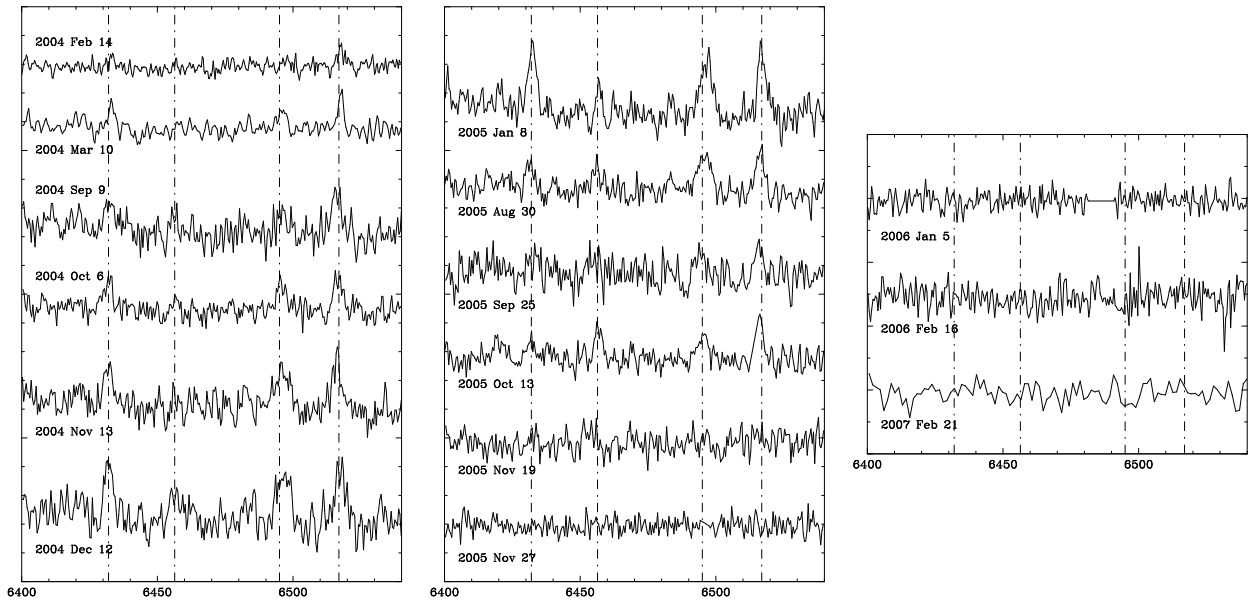} 
\caption{
The Fe lines in the 6400 to 6540~\AA\ region of the spectra.  These 
lines are [Fe~II] at 6432~\AA, Fe~II at 6457\AA, Fe~I at 6495~\AA, 
and Fe~II at 6517~\AA.  The vertical dot-dashed lines are at the 
line rest wavelengths.  The lines vary in intensity together, 
getting stronger from UT 2004 February 14 (when they are very weak) 
to UT 2005 January 8.  After this, they decline in intensity until 
UT 2005 November 27 when they are not detected.  Again, the spectra 
have been continuum subtracted and not smoothed.  
\label{fe-all}}
\end{figure}

\begin{figure}[tb] 
\epsscale{1.0}
\plotone{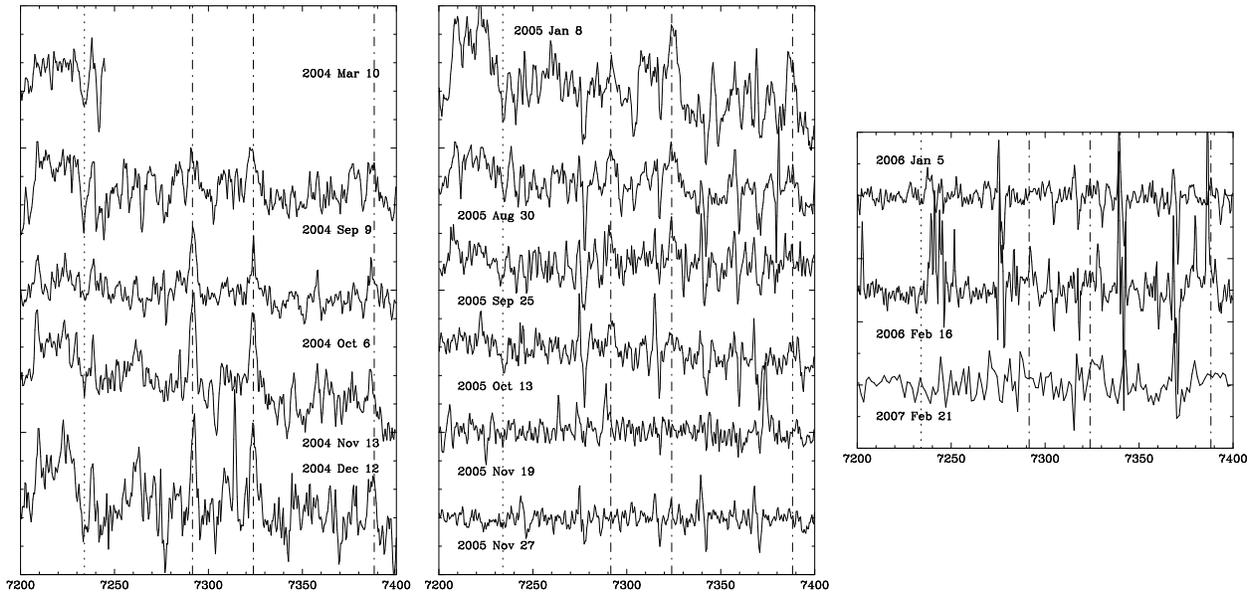} 
\caption{
The [Ca~II] lines in the 7200 to 7400~\AA\ region of the spectra.  
Specifically, at 7261, 7292, and 7324~\AA.  The vertical dot-dashed 
lines are at the line rest wavelengths.  At the long-wavelength 
edge of the plot is an [Fe~II] at 7388~\AA.  Our UT 2004 February 
14 and UT 2004 March 10 spectra did not cover this region of the 
spectra, however, from UT 2004 September 9 to UT 2005 October 13, 
the [Ca~II] lines are visible.  After this they are not present.  
Again, the spectra have been continuum subtracted and not smoothed.  
\label{caii-all}}
\end{figure}

The [O~I] emission line at 6300~\AA\ varied significantly in intensity over the monitoring period.  The changes are shown in Fig.~\ref{6300OI-all}.  We note that the wavelength of this line is coincident with a strong night-sky emission line although we consider sky subtraction to be generally very good.  At the start of the observing period the line was absent and appeared some time between 2004 February 14 and 2004 September 9 (unfortunately the spectrum from 2004 March 10 had a series of bad pixels at this wavelength and it was not possible to determine the presence of the line).  From 2004 September 9 onwards, the [O~I] line appeared to have an extended blue wing which persisted until the line faded.  The blue wing extended to $\sim$--270~km~s$^{-1}$ from the line rest wavelength and Fig.~\ref{6300OI-v} shows its profile from the spectrum taken on UT 2004 November 13.  The line continued to increase in intensity to 2005 January 8 where it was at a maximum.  Subsequently, the line faded but was still perhaps detectable in the low-resolution spectrum taken on 2007 February 21.  We note that the spectrum from 2006 February 16 had bad pixels at 6300~\AA\ which were removed by interpolation.  However, the wings of the [O~I] line are still detected.

The presence of forbidden emission lines in the optical spectra of young CTTSs has been studied by several authors (e.g. Edwards et al. 1987; Hartigan, Edwards, \& Ghandour 1995) and is generally associated with low-density outflowing gas.  Typically, only blueshifted emission wings are observed which has led to the model that the circumstellar disk occults redshifted emission from a latitude-dependent wind with the highest gas velocities at the polar region of the star.  From the models of Edwards et al. (1987), the only way to obtain [O~I] profiles like those seen in V1647~Ori, i.e. single-peaked with an extended blueshifted wing, is to view the wind from a low inclination angle, specifically,{\it i}$<$45$^{\circ}$ for a wind opening angle $\theta$ of $<$30$^{\circ}$ (see Fig.~10 in Edwards et al. 1987).  This upper limit to the inclination angle is somewhat at odds with that derived by Acosta-Pulido et al. (2004) and mentioned above.

Also, in Fig.~\ref{6300OI-all}, longward of the 6300~\AA\ [O~I] line, we observe an absorption feature that deepens and broadens.  Its wavelength is 6347\AA\ and we identify it as Si~II.  A weaker Si~II line is also seen at 6371~\AA.  The 6347~\AA\ line appears first weakly in the spectrum from 2004 March 10 and becomes stronger until 2005 January 8.  After this, it is possibly present until 2005 September 25 and then is absent.

\subsubsection{Fe emission lines}
In the wavelength range 6400 to 6540~\AA, there are four Fe emission lines from permitted (three) and forbidden (one) transitions.  These are [Fe~II] at 6432~\AA, Fe~II at 6457~\AA, Fe~I at 6495~\AA, and Fe~II at 6517~\AA.  The evolution of these lines is shown in Fig.~\ref{fe-all}.  The [Fe~II] 6432~\AA\ line is seen from the start of the monitoring period, 2004~February 14, to 2005 November 27 with peak emission occurring on 2005 January 8.  The Fe~II 6457~\AA\ line is relatively weak in all spectra from 2004 September 9 to 2005 November 27 and peak emission also occurs on 2005 January 8.  The Fe~I 6495~\AA\ line is seen from 2004 March 10 to 2005 October 13 and again, the emission peaks on 2005 January 8.  Finally, the Fe~II 6517~\AA\ line is visible from 2004 February 14 to 2005 November 19 and it is strongest on 2005 January 8.  The Fe~I 6495~\AA\ line is consistently broader than either the Fe~II or [Fe~II] lines and it has a full-width half maximum (FWHM) of 250~km~s$^{-1}$ compared to 150~km~s$^{-1}$ for the permitted and forbidden ionized Fe lines.  For comparison, the [Ca~II] line at 7292~\AA\ has a FWHM of 150~km~s$^{-1}$ while a CuAr arc lamp line located near the [Ca~II] line has a FWHM of 130~km~s$^{-1}$.

In Fig.~\ref{caii-all} we also identify an additional [Fe~II] emission line at 7388~\AA.  Its behavior is similar to that of the [Fe~II] 6517~\AA\ line although it disappears slightly earlier (2005 September 25 as opposed to 2005 November 27).  It also peaks in intensity on 2005 January 8.

The behavior of the [Fe~II] line at 6432~\AA\ seems consistent with that of the [S~II] lines described above and suggests a common excitation mechanism such as shock-excitation.  The permitted Fe lines are generally considered as indicators of chromospheric activity in T~Tauri stars (Beristain, Edwards, \& Kwan 1998) and therefore should show similar behavior to the [Ca~II] lines described below.

\subsubsection{[Ca~II] emission lines}
We have identified two [Ca~II] emission lines at 7292 and 7324~\AA.  The variations in these lines with time are plotted in Fig.~\ref{caii-all}.  Our 2004 February 14 and March 10 spectra did not include these lines, however, they are present in the spectra from 2004 September 9 to 2007 February 21.  The [Ca~II] lines are well detected from 2004 September 9 to 2005 August 30 and are weakly detected in the 2005 September 25 to 2005 October 13 spectra.  They are absent in the spectra taken after the latter date.  The [Ca~II] lines are unresolved with respect to the line width defined by the CuAr arc lamp lines (see above).

\subsubsection{The Keck~II HIRES spectrum}
The HIRES echelle spectra of V1647~Ori shows a number of both absorption and emission features with high-resolution providing information on their structure.  Below, we consider each spectral feature and discuss their appearance.

\begin{figure}[tb] 
\epsscale{0.8}
\plotone{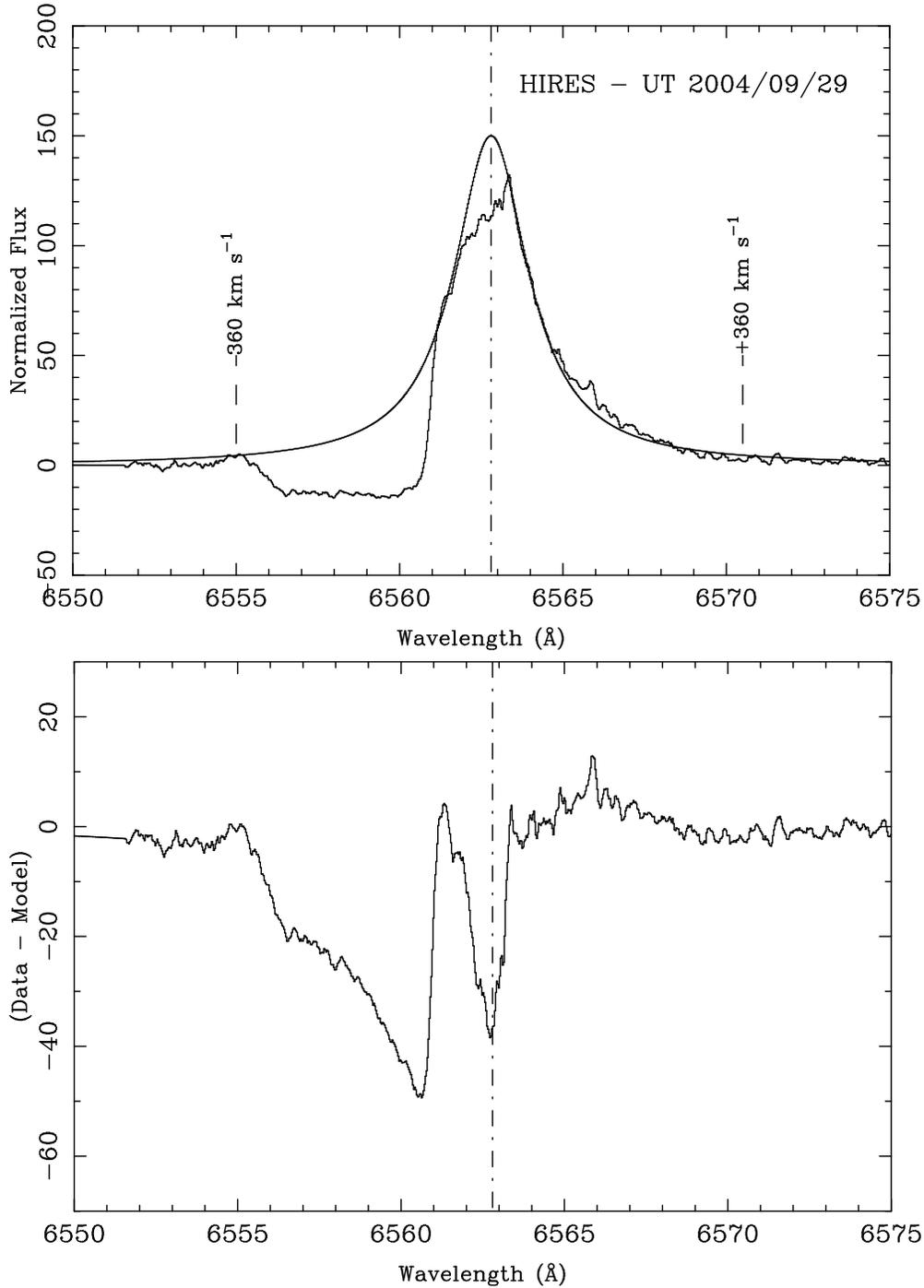} 
\caption{
The H$\alpha$ line profile from the HIRES spectrum of V1647~Ori.  
At the top is the observed spectrum overlaid with a model Gaussian
profile that fits the red wing of the line.  Note the wings of the 
profile extend to $\pm$360~km~s$^{-1}$ from the nominal wavelength 
of H$\alpha$ (dot-dashed line).  For comparison, an arc lamp line lying close to the wavelength of
H$\alpha$ has a full-width half-maximum gaussian profile of 12~km~s$^{-1}$.  
Below is the difference between 
the observed and model profile.  There appears to be significant 
H$\alpha$ absorption at zero velocity together with blueshifted 
absorption extending from  --80 to --360~km~s$^{-1}$.  The minimum 
in the H$\alpha$ absorption lies at 120~km~s$^{-1}$.
\label{hires-fit}}
\end{figure}

The H$\alpha$ line at high-spectral resolution is particularly interesting.  It has been previously observed at low-resolution to exhibit a P~Cygni profile with a strong blue-shifted absorption and very broad emission (e.g. Reipurth \& Aspin 2004; Walter et al. 2004).  Our HIRES spectra are the highest spectral resolution data (R$\sim$46,000) obtained on the source to date.  Fig.~\ref{hires-fit} shows the profile of H$\alpha$ from 2004 September 24, some 10 months after the eruption occurred.  At this time the source had a magnitude of r'$\sim$18.  The emission feature is asymmetric with an extended red wing reaching around +360~km~s$^{-1}$.  The blue-shifted absorption seems to begin close to the line rest wavelength and continues out to $\sim$--360~km~s$^{-1}$, the slight peak at this velocity appears to be where the blue wing of the emission line reaches the continuum level. This would make the emission symmetric about the line rest wavelength with a full-width zero intensity (FWZI) of around 720~km~s$^{-1}$.  The emission red wing appears to exponentially decay to the continuum.  It is difficult to say where the line emission peak flux is located due to the overlying absorption.  The absorption component appears to start close to the line rest wavelength and creates the very asymmetric profile seen in the blue wing.  This is probably formed by a combination of two absorption features.  The blue-shifted absorption goes below the local continuum level and is flat-bottomed from about --100 to --330~km~s$^{-1}$.  In Fig.~\ref{hires-fit} we show the H$\alpha$ profile together with a best-fit profile (top) fitting the red wing of the emission only.  Also shown (bottom) is the residual profile showing the difference between the observed and best-fit profile.  The profile of the fitted feature is Lorentzian (neither a Gaussian or Voigt profile fit the red wing well).  We note that the profile of an unresolved arc line (ThAr) is best-fit with a Gaussian of FWHM$\sim$6~km~s$^{-1}$.  The residual profile shown in Fig.~\ref{hires-fit} exhibits extensive absorption superimposed on the H$\alpha$ emission and clearly shows two distinct components, one at the H$\alpha$ rest wavelength (the dot-dashed line) with a FWHM of $\sim$50~km~s$^{-1}$, and the other extending from the continuum at --70~km~s$^{-1}$ out to --350~km~s$^{-1}$.  The more highly blue-shifted component has a more complex profile with the absorption being deepest at around --100~km~s$^{-1}$ then decreasing in an asymptotic manner to --290~km~s$^{-1}$ where the decrease becomes linear to the level of the continuum.

We detect the K~I 7665 and 7699~\AA\ lines which also possesses P~Cygni profiles.  The shorter wavelength line is located in the middle of the atmospheric O$_2$ absorption band, however, the P~Cygni structure is still evident and consistent with the longer wavelength line.  The P~Cygni absorption component has a velocity offset from the rest wavelength of --160~km~s$^{-1}$.  The 2004 October low-resolution GMOS spectrum shows these K~I lines which also display P~Cygni profiles.

The Na~D lines are present in absorption only and are broad, relatively symmetric, and blue-shifted by about --145~km~s$^{-1}$.  These are consistent with our low-resolution GMOS spectrum from our nearest date, UT 2004 October 6.  The signal to noise in the HIRES spectrum at the wavelength of the Na~D lines in insufficient (S/N~pixel$^{-1}\sim$5.5) to say if a narrow interstellar Na~D absorption component is present.

The 6347 and 6371~\AA\ Si~II lines are seen in absorption and both are shallow and broad (FWHM$\sim$--130~km~s$^{-1}$).  These lines are also seen in our low-resolution GMOS spectra from 2004 October 6.

\begin{figure}[tb] 
\epsscale{1.0}
\plotone{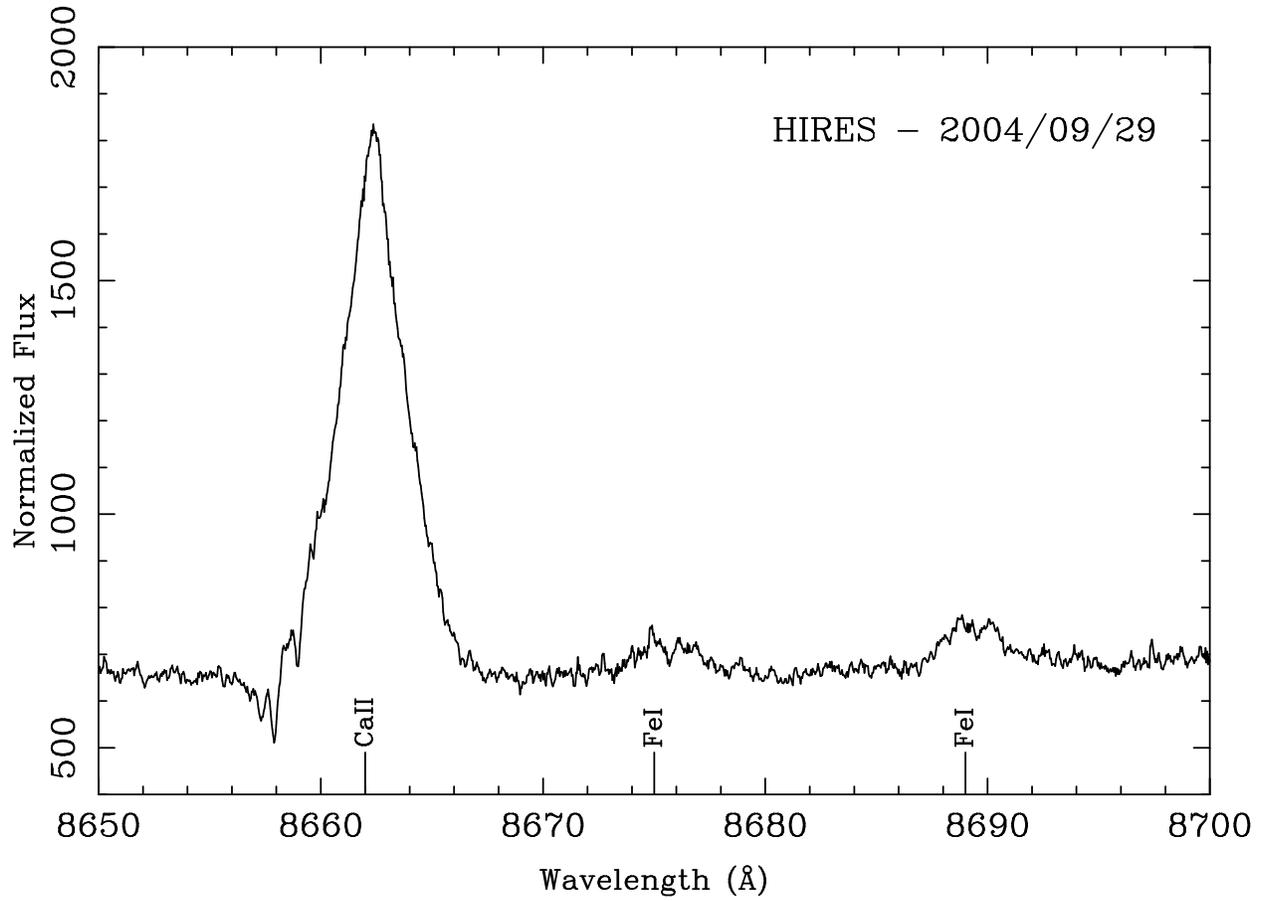} 
\caption{
The HIRES spectrum of V1647~Ori containing the Ca~II line at 8662\AA.  Note the trianglular profile of the line.  Also present are two faint Fe~I lines which seem to show a double-peaked nature.
\label{hires-caii}}
\end{figure}

The Ca~II line at 8498 and 8662~\AA\ lines are strongly in emission (their equivalent width are 10 and 8~\AA, respectively).  The middle of the Ca~II triple lines, at 8542~\AA, is unfortunately not within our observed spectral range.  The longer wavelength line has a pronounced blue-wing absorption feature extending to about --200~km~s$^{-1}$ from the line center.  The profile of this line is shown in Fig.~\ref{hires-caii}.  The line is triangular in profile as is the 8498~\AA\ line.

Strong line absorption at 7773~\AA\ corresponds to the O~I triplet lines.  This line is at the end of an echelle order and so no further information is obtainable.  In our low-resolution GMOS spectra, the O~I triplet is highly blended but is seen in absorption from 2004 September 3 to 2005 October 13.  After this the lines are either not present or hidden in the noise.

We detect two very weak Mg~II absorption lines at 7877 and 7897~\AA.  In our low-resolution spectrum from 2005 October 6, these lines are seen weakly in absorption confirming their detection in the HIRES data.

It seems that the 6300 and 6363~\AA\ [O~I] lines are very weakly in emission.  This is also seen in our 2005 October 6 low-resolution spectrum confirming their identification in the HIRES data.  The signal to noise of the HIRES detection is low but the lines appear real.  Also they are relatively narrow with respect to other emission lines with a FWHM of $\sim$40~km~s$^{-1}$.

A number of broad, weak emission lines of Fe~I (6 lines) and Ti~I (2 lines) are present.  The FWHM of both the Fe~I and Ti~I line is $\sim$100~km~s$^{-1}$.  There is a hint that the lines are double-peaked but, with the signal to noise present, it is difficult to say if this is correct or if they are broadened single lines or possibly overlapping pairs.

From the above detailed description of the HIRES spectra, we can conclude that the features present agree well with those seen in the 2004 October 6 low-resolution GMOS spectrum.  The HIRES spectrum also confirms that some of the weaker lines seen in the GMOS spectrum (with low signal to noise) are real.

\subsection{The variability of the H$\alpha$ line}
\begin{figure*}[tb] 
\epsscale{1.0}
\plotone{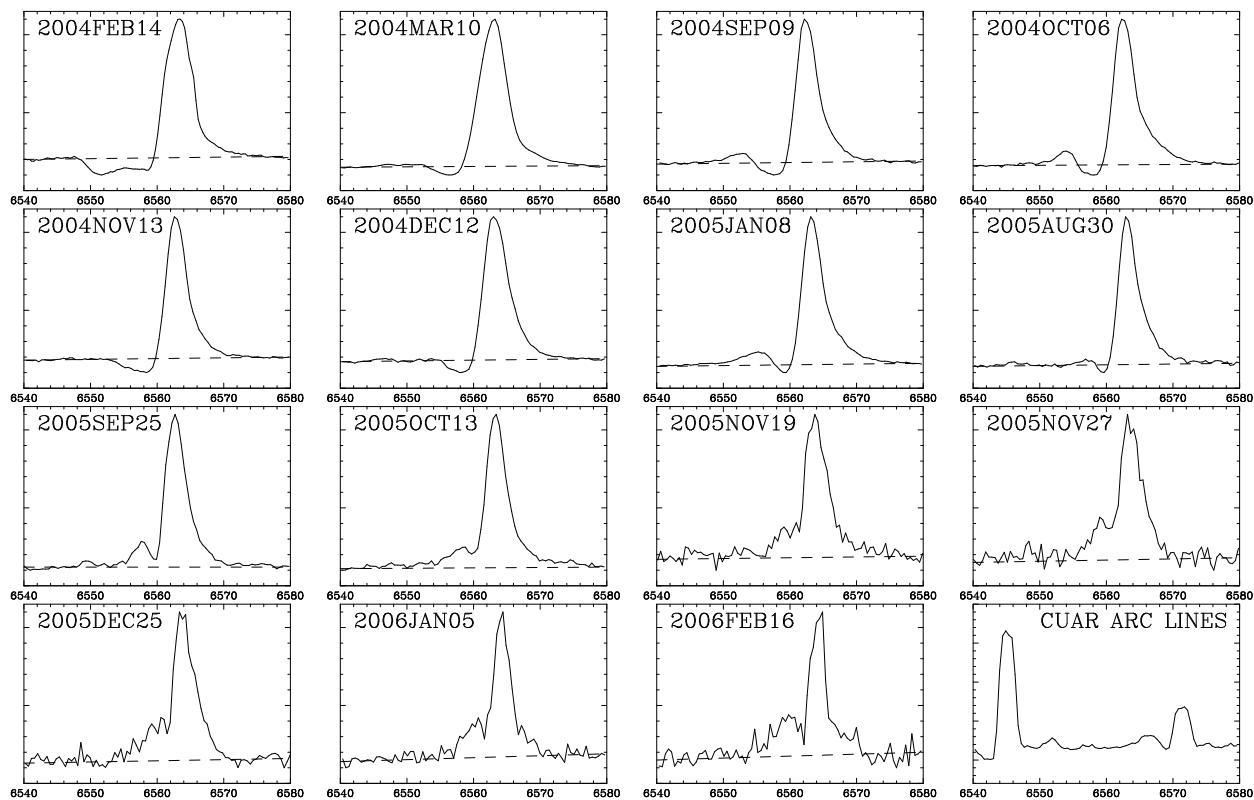} 
\caption{
H$\alpha$ emission line profiles extracted from the optical spectra of
V1647~Ori.  The monitoring period is 2004 February to 2006 February
and the date of observations is shown in the top-left corner.  To
better show the profile structure, the spectra have been scaled 
to a minimum and maximum flux of zero and unity, respectively.  The H$\alpha$
equivalent widths, and related information are given in Table~4.  Of
particular note is the prominent P~Cygni profile in the earlier
spectra.  The bottom-right plot is of CuAr arc lines for comparison of 
H$\alpha$ line widths and instrumental profile. 
\label{haplot1}}
\end{figure*}

\begin{figure*}[tb]
\epsscale{1.0} 
\plotone{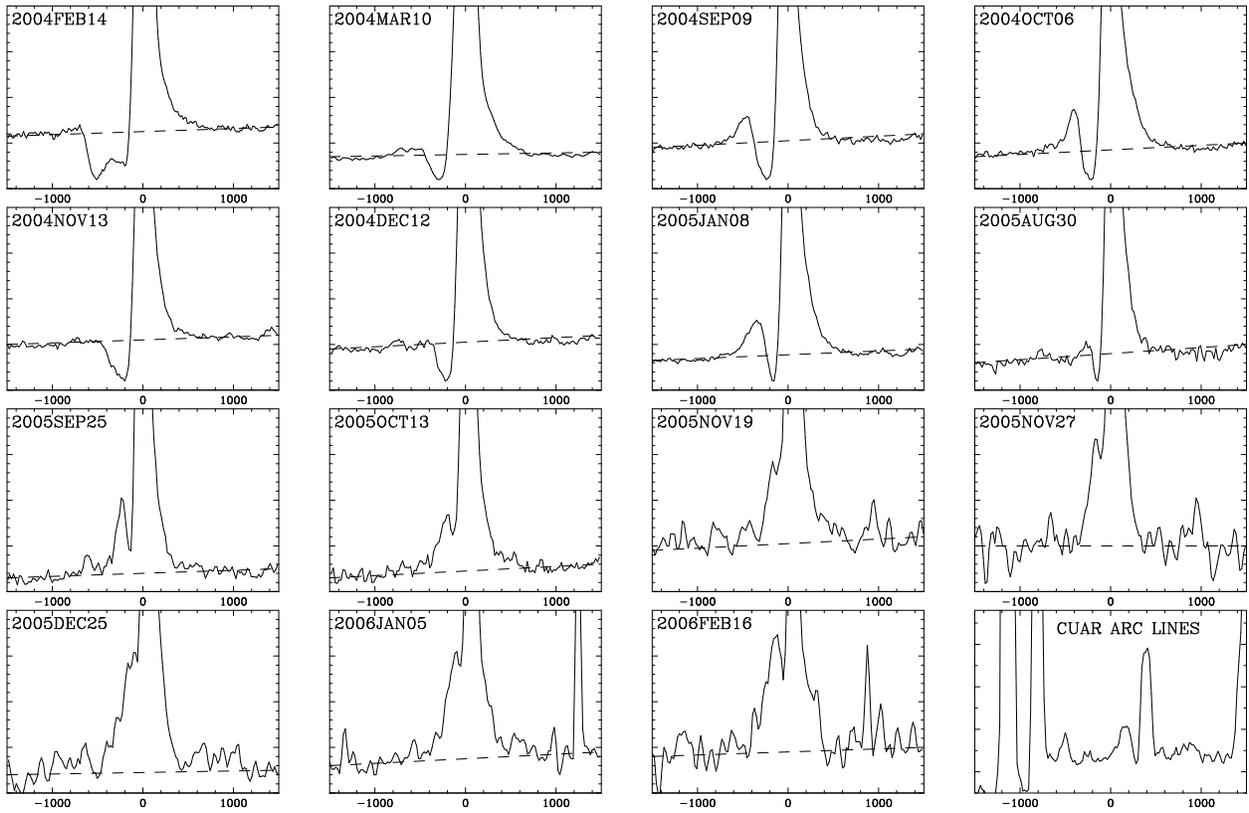}
\caption{ An expanded view of the H$\alpha$ P~Cygni profiles shown in
Fig.~\ref{haplot1}.  Here, the x-axis is velocity offset in
km~s$^{-1}$ from 6562.8~\AA.  Note the change in emission width,
absorption width and the trend for lower-velocity blue-shifted
absorption with increasing time.  
\label{haplot2}} 
\end{figure*}

In Fig.~\ref{haplot1}, we present the H$\alpha$ profile of V1647~Ori over the period February 2004 to February 2007.  These 15 observations show considerable variation in both emission and absorption characteristics as the eruption proceeded and during the decline phase.  The bottom right tile of the plot shows the instrumental profile (CuAr arc lamp lines) close to the wavelength of H$\alpha$. Fig.~\ref{haplot2} shows an expanded view of the H$\alpha$ profiles with an x-axis of velocity offset in km~s$^{-1}$ from the rest wavelength of H$\alpha$ (6562.8~\AA).

Qualitatively, the H$\alpha$ profile varies significantly even between consecutive observations. H$\alpha$ typically exhibits P~Cygni type structure with an emission component and a blue-shifted absorption feature. Throughout the outburst period, the emission component is extremely broad with a maximum FWZI of close to 1500~km~s$^{-1}$, however, it decreases in width as the eruption progresses. The width of the blue-shifted absorption is over 600~km~s$^{-1}$ at the start of the monitoring period and also becomes smaller with time.  Additionally, the absorption shifts towards the rest wavelength of H$\alpha$ as well as decreasing in strength. Throughout the first 18 months of the monitoring period, the absorption component absorbs below the continuum (see panels 2004FEB14 to 2005AUG30 in Figs.~\ref{haplot1} and \ref{haplot2}). After 2005AUG30, the wind absorption appears to be not strong enough to absorb below the continuum level and merely eats into the H$\alpha$ emission. Finally, we note that the H$\alpha$ emission and absorption components are not of sufficient signal-to-noise in our last three datasets (2005DEC25, 2006JAN05, and 2006FEB16) to reliably determine the extent of the absorption on the emission component.

We quantify the H$\alpha$ profile structure and variations in both Table~\ref{haprof} and Fig.~\ref{haplot3}.  In the latter, we show several measures of H$\alpha$ profile structure, specifically the equivalent width of the emission component (W$_{\lambda}$, top-right), the full-width zero intensity (I$_{max}$, FWZI) and the full-width at I$_{max}$/40 (2.5\%) intensity (FW2.5\%, middle-left), the velocity width of the blue-shifted absorption feature ($\Delta$V, middle-right), the velocity offset of the wavelength of deepest or 'characteristic' absorption (V$_{char}$, bottom-left), and the wavelength of peak (i.e. I$_{max}$) H$\alpha$ emission ($\lambda_{peak}$ bottom-right).  We show the FWZI and FW2.5\% variations since it is interesting to both compare the behavior of these measures of profile width and directly relate the latter to values seen in other stars, specifically from Reipurth, Pedrosa, \& Lago (1996, henceforth RPL96).  RPL96 pointed out that the FW2.5\% is considerably less sensitive to noise than FWZI.

From Fig.~\ref{haplot3} we see that as the eruption proceeded from early 2004 to early 2006 (detailed in the r' band brightness variations shown in the top-left panel), the behavior of the H$\alpha$ profile measures was rather complex.  Clearly, detailed numerical modeling of the profile changes is required to characterize and quantify the physical and geometric properties of the accretion and outflow during the eruption (see for example Kurosawa, Harries \& Symmington 2006).  However, we consider that the above five empirically defined quantities may give us some useful insight into the nature of the variability observed.

\subsubsection{H$\alpha$ equivalent width in the 'High Plateau' phase}
\begin{figure*}[tb]
\epsscale{0.9} 
\plotone{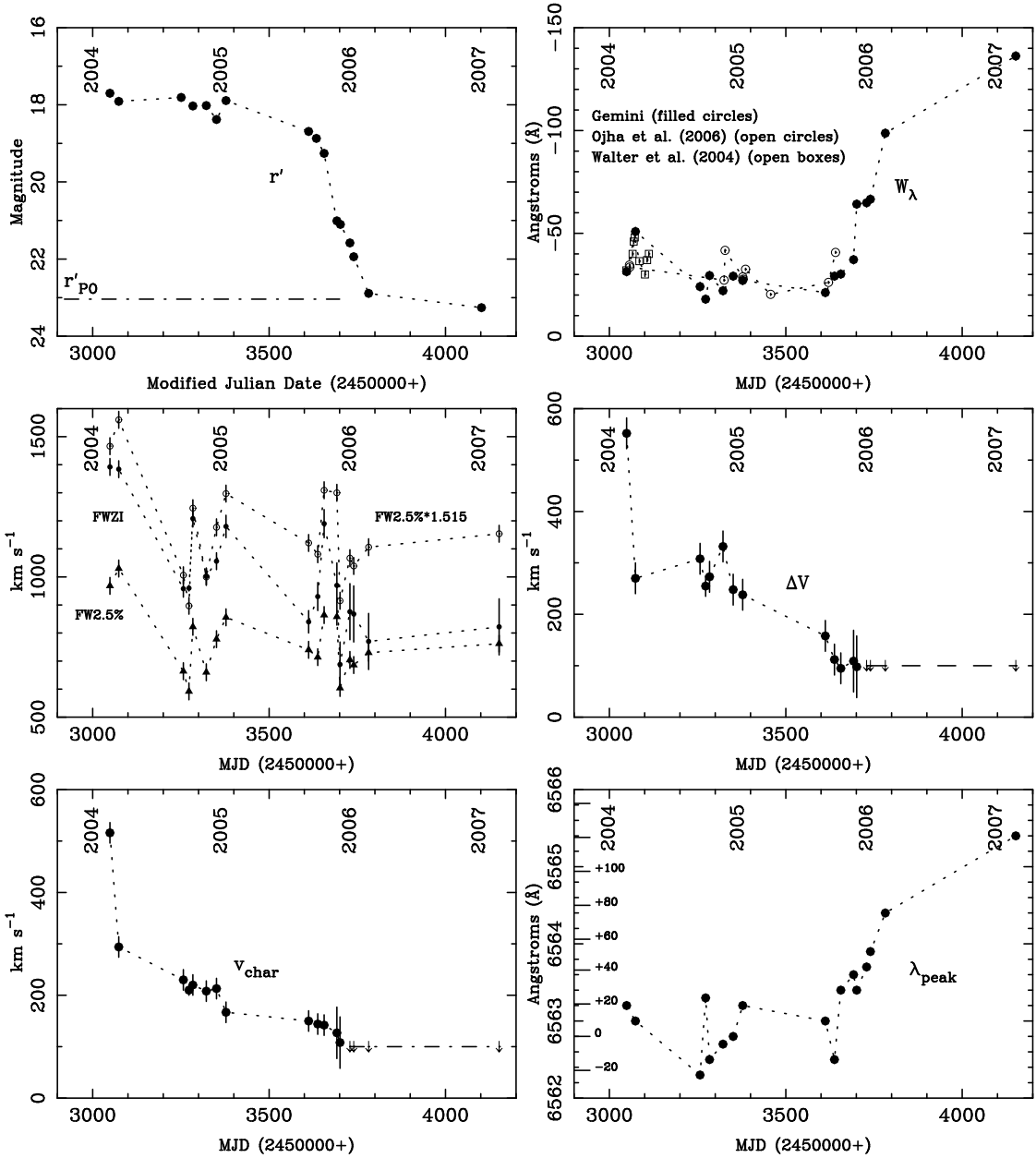}
\caption{ A plot of the changes in H$\alpha$ profile structure
  vs. time.    The units of the x-axes of all six panels are modified
  Julian Date (MJD).  The top-left panel shows the V1647~Ori r' band
  photometric light curve.  The horizontal dot-dashed line labeled
  r'$_{PO}$ is the pre-outburst SDSS r' brightness. The top-right
  panel shows the H$\alpha$ emission equivalent width (W$_{\lambda}$)
  in Angstroms.  The middle-left panel shows the Full Width Zero
  Intensity (FWZI), the Full Width 2.5\% Intensity (FW2.5\%), and the 
  scaled (see text for details) FW2.5\% Intensity of the emission component 
  in km~s$^{-1}$.  The
  middle-right panel shows the width of the blue-shifted absorption
  component ($\Delta$V) in km~s$^{-1}$.  The bottom-left panel
  shows the velocity offset (in km~s$^{-1}$ from 6562.8~\AA) of the deepest or 
  'characteristics' absorption (v$_{char}$). The bottom-right panel
  shows the wavelength of peak emission, termed $\lambda_{peak}$, in
  Angstroms. Velocity offsets (in km~s$^{-1}$ from 6562.8~\AA) are 
  given inside the y-axis.
  \label{haplot3}}
\end{figure*}

Our measurements of the variation of the H$\alpha$ emission W$_{H\alpha}$ vs. time are shown as the filled circles in the top-right panel of Fig.~\ref{haplot3}.  This demonstrates that from soon after the outburst began (2004 February) through to the start of the major decline phase (2005 September), W$_{H\alpha}$ was approximately constant, with perhaps a slight decline, and had a mean value around --30~\AA.  This time period was referred to as the 'high plateau' phase of the outburst decline by Acosta-Pulido et al. (2007) and we herein adopt this descriptive designation.  Over the high plateau period, the brightness of V1647~Ori showed a decline of about $\sim$1.2 magnitudes (see the top-left panel in Fig.~\ref{haplot3}).  This would result in an increase in W$_{\lambda}$ for constant H$\alpha$ emission flux.  Our average value of W$_{H\alpha}\sim$--30~\AA\ is consistent with the values presented in both Walter et al. (2004) and Ojha et al. (2006) whose data are also shown in Fig.~\ref{haplot3} (as open squares and open circles, respectively).  There are clearly short-term fluctuations in W$_{H\alpha}$ in all three datasets.  These variations were considered possibly periodic in nature by Walter et al. (2004) and Ojha et al. (2006).  Acosta-Pulido et al. (2007) detected short-term optical photometric variability in their light curves and, from a Fourier periodicity search, derived a 56 day period.  They commented that one possible interpretation of such a periodicity would be variable obscuration by circumstellar disk material orbiting the young star.

\subsubsection{Timescales of H$\alpha$ variability}
If such a periodicity is present in broad-band optical photometry or W$_{H\alpha}$, or both, then it is clearly an important phenomenon for investigating the physics occurring during the V1647~Ori outburst. In our spectroscopic data, we perhaps see some temporal structure that could be interpreted as periodic in nature. For example, in the period September 2004 to January 2005, the variations appear reasonably well phased with the data of Walter et al. (open squares). However, with so few observations (five) and their temporal spacing, we cannot confirm that such periodic variability exists. What we can do is characterize the shortest timescale upon which variability is found. To do this, we consider the minimum separation between consecutive observations (combining all data points) which show statistically significant variability and we adopt a $>$10\% change in W$_{H\alpha}$ as indicating statistical significance. There exists in the data only one date (MJD~3378) when observations were taken less than 1 day of each other (i.e. on the same night). Upon this date, the W$_{\lambda}$ values are consistent to better than 10\%. Next, there are three occurrences of consecutive observations separated by 1 day (i.e. taken on consecutive nights) and in all three cases variability was not detected at the $>$10\% level. The shortest time interval over which a $>$10\% change in W$_{H\alpha}$ is detected is 3 days. Specifically, between MJD~3066 and 3069, W$_{H\alpha}$ changed by +6~\AA\ (from $\sim$--40~\AA), a change of $\sim$15\%. We conclude that, from the limited data available, variability in W$_{H\alpha}$ is present in V1647~Ori on timescales as short as 3 days.

\subsubsection{H$\alpha$ equivalent width in the decline phase}
After the period of relatively constant W$_{H\alpha}$ (February 2004 to September 2005), W$_{H\alpha}$ begins to change dramatically as the major photometric decline phase starts. By the time the source has faded to close to its pre-outburst brightness (February 2006), W$_{H\alpha}$ has increased to about --100~\AA. The most obvious cause for such a dramatic change is the decrease in continuum emission as the source faded. From September 2005 to February 2006 the observed r' band magnitude decreased by $\sim$4 magnitudes or a factor 40. If the H$\alpha$ flux remained constant during this period then we would expect W$_{H\alpha}$ to have increased by a factor 40 from $\sim$--30~\AA\ to --1200~\AA. Since the increase in W$_{H\alpha}$ was measured to be only a factor $\sim$3.3, we can conclude that the H$\alpha$ emission flux must have significantly decreased from September 2005 to February 2006. We estimate this decline to be about a factor $\sim$12. If the H$\alpha$ emission is proportional to the mass accretion rate then this decline would correspond to an order of magnitude decrease in accretion from September 2005 to February 2006.  Using the relationship between near-IR H~I emission line flux and accretion rate defined by Muzerolle et al. (1998a), Acosta-Pulido et al. (2007) estimated that the mass accretion rate declined from $\sim$5$\times$10$^{-6}$~M$_{\odot}$~yr$^{-1}$ in March 2004 to $\sim$5$\times$10$^{-7}$~M$_{\odot}$~yr$^{-1}$ in May 2006. This factor 10 reduction in accretion rate is consistent with the 12$\times$ reduction in H$\alpha$ emission flux we observe between similar dates.

\subsubsection{Full Width Zero and 2.5\% Intensity}
The FWZI of H$\alpha$ emission can be considered a measure of twice the maximum velocity H$\alpha$ emitting gas attains in the process of line creation. However, we note that this quantity is not just affected by bulk gas motion, since line broadening can occur.  Muzerolle, Calvet, \& Hartmann (1998) suggested that Stark broadening could be an important mechanism in producing very large line widths.  Muzerolle et al. (2001) presented model H$\alpha$ line profiles under the influence of Stark broadening by considering a realistic range of physical parameters related to the accretion process, magnetosphere size, inclination angle, and stellar mass and temperature. A comparison of H$\alpha$ line profiles both with and without Stark broadening (and continuum opacity) showed that this effect can be a significant contributory factor with an increase of up to 2.5$\times$ in FWZI. Their consideration of such broadening resulted in line profiles that better matched those observed in T~Tauri stars. Stark broadening requires optically thick conditions and very high densities (n$_{H}<$10$^{12}$~cm$^{-3}$), both of which are found in post-shock regions of hot accretion shocks (Hartigan et al. 1991; Valenti, Basri, \& Johns 1993).

As mentioned above, one problem with studying FWZI is signal to noise (henceforth S:N). Poor S:N can mask the true wavelength at which line emission fades to the continuum level. In the case of V1647~Ori, this is particularly important in the later spectra when the source had faded below r'=20. Table~\ref{haprof} shows the S:N at both I$_{max}$ and in the adjacent continuum for all spectra. FWZI is likely a good measure of velocity width from February 2004 to the summer of 2005 where the S:N in the continuum is always $>$20. However, after this time the continuum S:N drops to $<$5. Alternatively, FW2.5\% of H$\alpha$ emission has been considered (by RPL96) to be a measure of line width less sensitive to S:N since it does not require a precise definition of continuum level. This is due to FW2.5\% being of a larger signal level and hence insensitive to continuum S:N issues.  RPL96 observed a significant number (63) of young low- and intermediate-mass stars at high spectral resolution (R$\sim$50000) and presented H$\alpha$ line profiles for each. The profiles were divided into four classes (Type I--IV) depending on the asymmetry of the line profile, and the location and depth of any associated absorption. The reader is referred to RPL96 for the detailed definitions of the classes.

The changes occurring in FWZI and FW2.5\% with time (middle-left panel of Fig.~\ref{haplot3}) mirror each other reasonable well from February 2004 to the middle of 2005. This suggests that our measurement of FWZI is most likely accurate. In the above time period, FW2.5\% is $\sim$1.4$\times$ smaller than FWZI. Subsequently, however, the FWZI and FW2.5\% values become much more similar suggesting that our values of FWZI are likely underestimates due to poor S:N and that the trend seen in FWZI is not purely intrinsic to the star. If we measure the FWZI and FW2.5\% of a pure gaussian profile we find that FW2.5\% is a factor 1.515 smaller than FWZI. Hence, the red wing of H$\alpha$ in V1647~Ori between February 2004 and mid-2005 have FWZI and FW2.5\% values of close to the gaussian ratio value. We have therefore multiplied the FW2.5\% values from August 2005 onwards by a factor 1.515 and these are displayed in Fig.~\ref{haplot3} (middle-left panel) as the FWZI values.

FW2.5\% decreases during 2004 from a maximum close to 1000~km~s$^{-1}$ (FWZI$\sim$1400~km~s$^{-1}$) to $\sim$750~km~s$^{-1}$ (FWZI$\sim$1100~km~s$^{-1}$). After this time, the decrease in FW2.5\% ceases and from 2005 onwards it remains at $\sim$750~km~s$^{-1}$ (scaled FWZI$\sim$1150~km~s$^{-1}$). Hence, through the major decline phase of V1647~Ori until it reaches its pre-outburst brightness, FW2.5\% remains approximately constant. We note that the values encountered in V1647~Ori throughout the eruption are considerably larger (by a factor $\sim$3$\times$) than those found in CTTSs suggesting that accretion activity has not declined completely to a quiescent level.

The range of values of FW2.5\% seen in V1647~Ori is 750--1500~km~s$^{-1}$.  Fig.~9 of RPL96 shows that of their 63 sources, the peak in the distribution of (red wing) FW2.5\% values lies at $\sim$600~km~s$^{-1}$ with only three stars showing values greater than 1000~km~s$^{-1}$.  Hence, even when V1647~Ori had faded to its pre-outburst brightness, its FW2.5\% value remained larger than the RPL96 distribution peak.  The three sources that RPL96 found to have FW2.5\%$>$1000~km~s$^{-1}$ were R~Mon, FW~Cha, and AS~353A.  These sources were all classified as III-B\footnotemark\footnotetext{Double-peaked H$\alpha$ emission with
  the peaks being of unequal height and the secondary peak being $<$50\%
  of the primary peak} in the RPL96 scheme possessing blue-shifted absorption.  In total, 15 of their 63 sources showed type III-B profiles.  The H$\alpha$ profiles of V1647~Ori suggests that its RPL96 classification evolved with time from i) IV-B\footnotemark\footnotetext{P~Cygni profile with absorption at
  sufficiently high velocity absorption to absorb continuum flux
  beyond the wing of the H$\alpha$ emission.} in February/March 2004, to ii) III-B in September/October 2004, to iii) IV-B in November/December 2004, to iv) III-B in January 2005 through to at least October 2005 and possibly beyond to February 2007 (although the S:N of the spectra are too poor to be definite).

\subsubsection{Absorption width and velocity offset}
The width of the blue-shifted absorption feature (herein termed $\Delta$V) and the velocity of the center of the blue-shifted absorption (herein termed v$_{char}$) gives us information on the velocity field of the outflowing absorbing gas. Specifically, we consider that we can associate $\Delta$V with the velocity dispersion in the ejected material, and v$_{char}$ with its 'characteristic' velocity.

The temporal variability of $\Delta$V (middle-right panel of Fig.~\ref{haplot3}) is considerably different from that of the H$\alpha$ emission component. A very rapid decline in absorption width is seen between our first two observations taken in February 2004 ($\Delta$V$\sim$600~km~s$^{-1}$) and March 2004 ($\Delta$V$\sim$300~km~s$^{-1}$). After this time, $\Delta$V is approximately constant to the end of our winter 2004--2005 monitoring season. At the start of the fall 2005 observing period, $\Delta$V has decreased to $\sim$100~km~s$^{-1}$ and there is some suggestion that this decrease was related to the main photometric decline. During the fall/winter months of 2005, $\Delta$V remains approximately constant although the S:N of the spectra is poor. The final observation point, February 2007, also had very low S:N but the indication is that the value of $\Delta$V had changed little.

The velocity offset of the deepest absorption, v$_{char}$ (bottom-left panel, dashed line), mirrors $\Delta$V showing a similar large change between the first two observations (from $\sim$500~km~s$^{-1}$ to $\sim$300~km~s$^{-1}$). This is followed by a slower decline (to $\sim$200~km~s$^{-1}$) between the spring and fall of 2004. In the fall of 2005 v$_{char}$ had again decreased, this time by a factor $\sim$2 (to $\leq$100~km~s$^{-1}$). By early 2006, when V1647~Ori had returned to its pre-outburst brightness, the S:N of the spectra and the weakness of the absorption made a determination of v$_{char}$ difficult and all we can say is that it was $\leq$100~km~s$^{-1}$.

\subsubsection{Peak emission wavelength}
The wavelength of peak H$\alpha$ emission is a rather complex quantity since it is dependent not only on the motion of the emitting material, but also on the characteristics of any associated absorption.  For example, if a blue-shifted wind absorbs the blue wing of the emission line then the wavelength of peak emission can shift to the red.  Alternatively, if the blue-shifted absorption is weak and we measure a wavelength shift in peak emission, then we could interpret this as evidence for bulk motion of the emitting gas.  Multiple absorption components, as we saw in Fig.~\ref{hires-fit}, also complicates the interpretation.  In light of the above discussion, we refrain from any ad hoc interpretation of the changes in $\lambda_{peak}$ and merely relate the variations observed.

The final panel of Fig.~\ref{haplot3} shows the wavelength of peak H$\alpha$ emission ($\lambda_{peak}$, bottom-right panel) over the monitoring period. Here we see a peak-to-peak variability of $\sim$3~\AA\ ($\sim$140~km~s$^{-1}$) from February 2004 to February 2007. Soon after outburst ( February/March 2004), the peak in emission was measured to be slightly to the red of our assumed H$\alpha$ rest wavelength (6562.8~\AA). By the fall of 2004, $\lambda_{peak}$ had shifted to the blue by about 1~\AA\ ($\sim$45~km~s$^{-1}$). During the fall/winter of 2004, the peak had shifted back to its original wavelength (of February/March 2004). When our monitoring commenced again in the fall of 2005, $\lambda_{peak}$ was at about the same wavelength as in spring 2005 but soon after it shifted briefly to the blue and then rapidly to the red and by the spring of 2006 it was located redward of the rest wavelength by $\sim$1.7~\AA\ ($\sim$80~km~s$^{-1}$). This trend continued until our last observation, in February 2007, when $\lambda_{peak}$ was at +2.5~\AA\ ($\sim$120~km~s$^{-1}$).

\subsection{H$\alpha$ Profile Modeling}
It is generally agreed that the presence of complex H$\alpha$ emission structure with blue-shifted absorption in the spectrum of a young star is the result of the combination of emission from magnetospheric accretion and absorption from strong, dense stellar winds.  The recent papers by Muzerolle, Calvet \& Hartmann (1998), Alencar et al. (2005), Kurosawa, Harries, \& Symington (2006), and Bouvier et al. (2007) discuss models involving such processes and their fit to observations of low-mass stars.  It is clearly important to consider such numerical simulations to available datasets to allow the derivation of physical parameters such as outflow and accretion rates, and test the validity of the specific models.  However, due to the complexity of such analyses in regions of high accretion (Kurosawa, private communication), what follows will concentrate on a qualitative understanding of the changes observed.

\section{Physical Components of the Outburst}
The eruptive event that commenced between 2003 October 23 and 2003 November 15 (Brice\~no et al. 2004) resulted in the dramatic brightening of V1647~Ori and the appearance of what is now known as McNeil's Nebula. The event was apparently extremely violent since about three months after the outburst occurred (when our first observations were taken) the star had optically brightened by a factor $\sim$100 and produced a P~Cygni H$\alpha$ profile exhibiting both a very high velocity ($\sim$600~km~s$^{-1}$), blue-shifted, absorption trough (indicative of the presence of a dense wind) {\it~and} strong emission with an enormous velocity width (i.e.  FWZI$\sim$1500~km~s$^{-1}$ - indicative of magnetospheric accretion and perhaps Stark broadening). How these H$\alpha$ spectral features evolved over the first three months of the eruption is unknown, however, they are likely to be at least similar in nature to those observed at three months old if not even more extreme. Below, we summarize what we can deduce from our results regarding the properties and evolution of the wind and accretion phenomena.

\subsection{The Fast Wind}
At the three month mark of the eruption, a strong wind existed producing two minima in the blue-shifted absorption (see Fig.~\ref{haplot2}). Such a double-trough absorption profile was evident at times, for example, in the spectra of V1057~Cyg, a classical FUor (see Bastian \& Mundt 1985). However, the two components of the wind disappeared between our first and second observations on UT 2004 February 14 and UT 2004 March 10, respectively. The optical spectrum of Ojha et al. (2005) from UT 2004 February 22 did not show such features either. The two minima in the H$\alpha$ absorption on UT 2004 February 14, were at velocities of v= --270, and --530~km~s$^{-1}$ with respect to the rest wavelength of H$\alpha$ (6562.8~$\mu$m). On UT 2004 March 10, the single absorption minimum was at v=~--350~km~s$^{-1}$. A double-peaked absorption could be associated with multiple shell-like wind components ejected with different velocities or with the geometry and motion of the emission region. A continuing slowing of the wind is indicated by the v$_{char}$ values plotted in Fig.~\ref{haplot3} where the velocity of the absorption minimum declined from a maximum of --530~km~s$^{-1}$ to a minimum of --150~km~s$^{-1}$. The width of the absorption component, $\Delta$V, also declined over this period (from 560 to 100~km~s$^{-1}$) indicating a reduction in velocity dispersion in the wind and suggesting that the wind decelerated as it expanded away from V1647~Ori. Such behavior is perhaps consistent with the stochastic wind models of Grinin \& Mitskevich (1992) and Mitskevich, Natta, \& Grinin (1993). In these models the wind is non-isotropic and clumpy and decelerates at large distances from the young star.

\subsection{The Emission Region}
During the period of rapid decline of v$_{char}$, i.e. between 2004 February 14 and March 10, the FWZI of the H$\alpha$ emission remained relatively constant at $\sim$1500~km~s$^{-1}$.  As we saw above, such a value is somewhat larger than typically found in accreting classical T~Tauri stars (from the work of RPL96).  Additional examples of this are RW~Aur, which exhibited a FWZI of $\sim$1000~km~s$^{-1}$ in the spectra of Alencar et al. (2005), and some sources studied by Muzerolle, Hartmann, \& Calvet (1998) and Alencar \& Basri (2000) where a typical FWZI was found to be in the range 600 to 1000~km~s$^{-1}$.  The FWZI did not show significant decline until sometime between our observation in March 2004 and our next observation in September 2004.  Over this period the FWZI had reduced by around 500~km~s$^{-1}$ (to $\sim$1000~km~s$^{-1}$).  During the period of the major photometric decline, from August 2005 to February 2006, the FWZI remained in the range 1000--1400~km~s$^{-1}$, i.e. the dramatic change in optical brightness of V1647~Ori did not appear to be strongly correlated with the FWZI of the H$\alpha$ emission.  This argues that the optical fading and the H$\alpha$ emission are not the result of the same physical processes.

As we related above, contributions to the FWZI of the H$\alpha$ emission probably arise in both bulk gas motion and Stark broadening.  The region of H$\alpha$ emission therefore much be extremely dense and suggests that it is the post-shock region of the accretion shock. The change in optical colors as the major photometric decline phase began (they became redder) suggests that there was a reduction in blue continuum from the accretion shock region. Brice\~no et al. (2004) observed the spectrum of V1647~Ori using reflected light from McNeil's Nebula and found it to correspond to that from an early B-type source i.e. the hot accretion shock. Correspondingly, the reduction in blue continuum likely resulted from of a decline in post-shock temperature and density, probably due to a slowing of the accretion rate. The decline in accretion rate was estimated, using near-IR H lines, to be an order of magnitude between March 2004 and May 2006 (Acosta-Pulido et al. (2007).

\section{Outburst Timeline}
We now attempt to piece together the timeline of events over the outburst period and relate the photometric and spectroscopic variations discussed above.

\begin{enumerate}
\item Prior to the outburst, V1647~Ori appeared to be in an unstable
  state with variations in optical brightness of up to $\sim$2
  magnitudes.  It appeared that the main outburst event was initiated
  from an photometrically elevated state existing for at least 5 years
  prior to the main eruption i.e. from the first Brice\~no et
  al. (2004) and McGehee et al. (2004, SDSS) observations.

\item The outburst began between late-July and late-October 2003.
  Unfortunately, there are only a handful of photometric measurements
  for the first three+ months of the event.  These were obtained
  serendipitously during an Orion~OB1 photometric monitoring program
  (Brice\~no et al. 2001).  The rise time to peak brightness was
  short, 120 days, with a change in optical (I-band) brightness of 3.6
  magnitudes.\footnotemark\footnotetext{V1647~Ori had already
    brightened by $\sim$1 magnitude prior to this but it is unknown
    when this occurred due to lack of observations during this period
    (Brice\~no et al. 2004).}

\item At the time of our first observations, the first taken on the
  source after its discovery by McNeil (2004), the star had brightened
  from its pre-outburst level by over 5 magnitudes in the optical and
  over 2 magnitudes at 2~$\mu$m.  It had optical spectral features
  indicative of a massive accretion burst producing a fast, dense
  stellar wind creating high-velocity blue-shifted P~Cygni-type
  absorption on strong H$\alpha$ emission.  In the NIR, V1647~Ori
  possessed strong emission in Br$\gamma$, Na~I, and the $v=2-0$ CO
  overtone bandheads, all indications were that the emission region
  was hot, extremely dense, and in an excited kinematic state.  The
  optical brightening is attributed to a combination of the star being
  revealed after dust in the immediate circumstellar environment of
  the object was sublimated (Aspin et al. 2009) together with a
  luminosity increase due to the addition of significant hot accretion
  flux.  The brightening in the NIR is perhaps more likely the effect
  of irradiation and subsequent heating of the inner regions of the
  circumstellar disk, beyond the sublimation radius, by UV flux from
  the accretion flow.  However, it is clear that the heating did not
  propagate through to outer regions of the disk since the sub-mm/mm
  flux remained unchanged.

\item Over the next 10 months, V1647~Ori remained within half a
  magnitude of its outburst optical/NIR brightness although there was
  seemingly stochastic variability at the level of around 1 magnitude
  peak-to-peak.  Statistically significant brightness fluctuations
  were seen on periods as short as a few days.  During this so-called
  ``high-plateau'' phase, the optical spectrum of V1647~Ori showed
  strong P~Cygni H$\alpha$ emission on a relatively stable continuum.
  This suggests that the accretion process continued unabated over
  this period with only minor variations occurring.  This is supported
  by the reasonably constant FWZI of H$\alpha$
  ($\sim$1500~km~s$^{-1}$), indicating that the emission was taking
  place in very dense gas, and the derived accretion rate from NIR
  line emission.  During this same period, the wind quickly slowed
  (from v$_{char}\sim$600~km~s$^{-1}$ to 200~km~s$^{-1}$).  Also, the
  velocity dispersion in the wind mirrored v$_{char}$ in that it
  quickly declined (from $\Delta$V$\sim$600~km~s$^{-1}$ to
  200~km~s$^{-1}$).  During the rapid decline in wind velocity and
  velocity dispersion, weak shock-excited emission lines of [S~II]
  appeared in the optical spectrum.  These remained approximately
  constant to the end of the monitoring period.  During the summer
  months of 2004, an [O~I] emission line appeared and strengthened for
  about a year.  This may be the result of the formation of a partly
  obscured bipolar-like outflow and its formation approximately
  corresponded to the period of the high-plateau.  Subsequently, the
  [O~I] line declined in intensity which corresponded to the rapid
  decline in brightness of V1647~Ori.

\item At the start of the rapid decline phase of the optical flux (in
  the fall of 2005, some 6 months after our last plateau-phase
  observation), V1647~Ori had already shown a significant reduction in
  H$\alpha$ FWZI ($\sim$1500~km~s$^{-1}$ to $\sim$800~km~s$^{-1}$),
  wind velocity (v$_{char}$$\sim$300~km~s$^{-1}$ to
  $\sim$200~km~s$^{-1}$), wind velocity dispersion
  ($\Delta$V$\sim$250~km~s$^{-1}$ to $\sim$150~km~s$^{-1}$), and
  accretion rate (a factor 10). This suggests that the
  density of the region of H$\alpha$ emission declined and the wind
  slowed.  Two of the three aforementioned parameters (FWZI
  and $\Delta$V) remained approximately constant to the end of our
  monitoring period (in 2006 February) when the source had faded by
  $\sim$4 magnitudes and returned to its pre-outburst optical
  brightness.  The only significant change was in wind velocity which
  continued to decline ($\delta$v$\sim$250~km~s$^{-1}$ to
  $\sim$100~km~s$^{-1}$) into early 2006 implying that the wind was
  still decelerating.

\item In the period after February 2006, beyond the time period
  investigated here, we obtained one further set of observations that
  proved rather interesting.  These were published in Aspin, Beck, \&
  Reipurth (2008) and were obtained approximately one year after the
  last observation presented above.  In February 2007, the optical
  brightness of V1647~Ori remained close to its February 2006 value
  with r'=23.3.  Little evidence was seen for blueshifted H$\alpha$
  absorption suggesting that the wind had continued to weaken over the
  intervening year.  The FWZI of H$\alpha$ remained at
  $\sim$800~km~s$^{-1}$ implying that the accretion process had
  perhaps settled into a quasi-steady state.  The H$\alpha$ line
  profile calculations by Muzerolle, Calvet, \& Hartmann (2001)
  discussed above, gave values of FWZI of this order (see their
  Figs.~8, 11, and 12) and produced a good match to observations of
  some CTTSs (e.g. BP~Tau with FWZI$\sim$800~km~s$^{-1}$).  In
  addition, the February 2007 data showed that shock-excited [S~II]
  emission was still present, as was weak [O~I] emission with a weak
  blueshifted component.

\end{enumerate}

\vspace{0.3cm}
{\bf Acknowledgments} 

We are grateful to G.H.~Herbig for providing the HIRES spectrum of V1647~Ori presented above.  We also thank C.~Brice\~no for access to his 2004 February 18 spectrum of V1647~Ori.  We additionally thank D.~Finkbeiner and P.~McGehee for assistance with accessing SDSS data, and the latter for providing numerical model information.  This work is based on observations obtained at the Gemini Observatory (under program identifications GN-2004A-DD-3, GN-2004B-Q-28, GN-2005B-Q-1), which is operated by the Association of Universities for Research in Astronomy, Inc., under a cooperative agreement with the NSF on behalf of the Gemini partnership: the National Science Foundation (United States), the Particle Physics and Astronomy Research Council (United Kingdom), the National Research Council (Canada), CONICYT (Chile), the Australian Research Council (Australia), CNPq (Brazil) and CONICET (Argentina).  Funding for the SDSS and SDSS-II has been provided by the Alfred P. Sloan Foundation, the Participating Institutions, the National Science Foundation, the U.S. Department of Energy, the National Aeronautics and Space Administration, the Japanese Monbukagakusho, the Max Planck Society, and the Higher Education Funding Council for England. The SDSS Web Site is http://www.sdss.org/.  The SDSS is managed by the Astrophysical Research Consortium for the Participating Institutions.  The Participating Institutions are the American Museum of Natural History, Astrophysical Institute Potsdam, University of Basel, Cambridge University, Case Western Reserve University, University of Chicago, Drexel University, Fermilab, the Institute for Advanced Study, the Japan Participation Group, Johns Hopkins University, the Joint Institute for Nuclear Astrophysics, the Kavli Institute for Particle Astrophysics and Cosmology, the Korean Scientist Group, the Chinese Academy of Sciences (LAMOST), Los Alamos National Laboratory, the Max-Planck-Institute for Astronomy (MPIA), the Max-Planck-Institute for Astrophysics (MPA), New Mexico State University, Ohio State University, University of Pittsburgh, University of Portsmouth, Princeton University, the United States Naval Observatory, and the University of Washington.  This research has made use of NASA's Astrophysics Data System Bibliographic Services. This research has made use of the SIMBAD database, operated at CDS, Strasbourg, France.  We are most fortunate to have the opportunity to conduct astronomical observations from the sacred mountain Mauna Kea.  During this research, CA was supported in part by NASA through the American Astronomical Society's Small Research Grant Program.  BR acknowledges partial support from the NASA Astrobiology Institute under Cooperative Agreement No. NNA04CC08A.

% REFERENCES 

\clearpage 

%############################################################################################
% Table 1
\begin{deluxetable}{ccllr}
\tabletypesize{\scriptsize}
\tablecaption{V1647~Ori Gemini GMOS-N Observing Log\label{obslog}}
\tablewidth{0pc}
\tablehead{
\colhead{UT Date} & \colhead{MJD\tablenotemark{a}} & \colhead{Filters and/or} & \colhead{Exposure Times} & \colhead{Seeing} \\  
\colhead{yymmdd}& \colhead{}                     & \colhead{Grism Used}     & \colhead{(seconds)}        & \colhead{(r'-band)} }

\startdata
981116  & 1134 & g,r,i,z\tablenotemark{b}& --          & 0$\farcs$9 \\
040214  & 3049 & g',r',i',z'\tablenotemark{c},R831\tablenotemark{d} & 60,60,60,60 & 0$\farcs$54 \\
040310  & 3074 & r',B600\tablenotemark{e}                           & 10,1200          & 0$\farcs$96 \\
040903  & 3251 & g',r',i',z'        & 30,30,30,30      & 0$\farcs$53 \\
040909  & 3257 & B600               & 900              & 0$\farcs$62 \\
040924  & 3272 & HIRES              & 3600             & ?           \\
041006  & 3284 & g',r',i',z',B600   & 30,30,30,30,900  & 0$\farcs$48 \\
041113  & 3322 & g',r',i',z',B600   & 30,30,30,30,1200 & 0$\farcs$49 \\
041212  & 3351 & g',r',i',z',B600   & 30,30,30,30,1200 & 0$\farcs$58 \\
050108  & 3378 & g',r',i',z',B600   & 30,30,30,30,1200 & 0$\farcs$78 \\
050830  & 3612 & g',r',i',z',B600   & 30,30,30,30,900  & 0$\farcs$63 \\
050921  & 3634 & g',r',i',z'        & 30,30,30,30      & 0$\farcs$48 \\
050925  & 3638 & B600               & 900              & 0$\farcs$65 \\
051013  & 3656 & g',r',i',z',B600   & 30,30,30,30,900  & 0$\farcs$68 \\
051119  & 3692 & r',B600            & 30,900           & 0$\farcs$60 \\
051127  & 3701 & B600               & 30,1200          & 0$\farcs$61 \\
051128  & 3702 & g',r',i',z'        & 60,60,60,60      & 0$\farcs$59 \\
051225  & 3729 & g',r',i',z',B600   & 60,60,60,60,1200 & 0$\farcs$59 \\
060105  & 3740 & g',r',i',z',B600   & 60,60,60,60,1200 & 0$\farcs$49 \\
060216  & 3782 & r',B600            & 120,3600         & 0$\farcs$42 \\
061222\tablenotemark{f} & 4090 & R  & 30               & 0$\farcs$90 \\
070221  & 4152 & R400               & 3.5 hours        & 0$\farcs$76 \\
070222  & 4153 & g',r',i',z'        & 600              & 0$\farcs$60 \\
\enddata

\tablenotetext{a}{Modified Julian Date. 2450000+}
\tablenotetext{b}{SDSS observations taken from the galactic plane Orion release data (Finkbeiner et al. 2004).}
\tablenotetext{c}{GMOS ``Sloan'' filters.}
\tablenotetext{d}{GMOS Red 831 lines/mm grating, blaze wavelength 757nm, R$\sim$4396 with 0$\farcs$5 slit,
simultaneous wavelength coverage 207nm, 0.034 nm/pixel.}
\tablenotetext{e}{GMOS Blue 600 lines/mm grating, blaze wavelength 461nm, R$\sim$1688 with 0$\farcs$5 slit,
simultaneous wavelength coverage 276nm, 0.045 nm/pixel.}
\tablenotetext{f}{Taken on University of Hawaii 2.2m telescope.}

\end{deluxetable}
\clearpage
%############################################################################################
% Table 2
\begin{center}
\begin{deluxetable}{ccccccc}
%\tabletypesize{\scriptsize}
{\scriptsize}
\tablecaption{V1647 Ori and comparison star coordinates and photometry\tablenotemark{a}\label{calstars}}
\tablewidth{0pc}
\rotate
\tablehead{
\colhead{Source} & \colhead{R.A.}    & \colhead{Decl.}   & \colhead{ g'} 
& \colhead{r'} & \colhead{i'} & \colhead{z'} \\  
\colhead{}       & \colhead{(J2000)} & \colhead{(J2000)} & \colhead{(Mags)} 
& \colhead{(Mags)}               & \colhead{(Mags)}               & \colhead{(Mags)} }

\startdata
V1647 Ori & 05 46 13.1 & -00 06 05 & 24.74$\pm$0.76   & 23.04$\pm$0.22 & 20.81$\pm$0.05 & 18.80$\pm$0.04 \\
S1        & 05 46 13.0 & -00 08 15 & 23.09$\pm$0.15   & 21.14$\pm$0.04 & 19.47$\pm$0.02 & 18.37$\pm$0.03 \\
S2        & 05 46 12.3 & -00 08 08 & 23.29$\pm$0.18   & 21.19$\pm$0.05 & 18.74$\pm$0.02 & 16.93$\pm$0.02 \\
S3        & 05 46 11.4 & -00 07 55 & 23.26$\pm$0.17   & 21.39$\pm$0.05 & 20.71$\pm$0.05 & 19.65$\pm$0.07 \\
S4        & 05 46 16.3 & -00 06 52 & 21.93$\pm$0.06   & 20.30$\pm$0.02 & 18.58$\pm$0.02 & 17.54$\pm$0.02\\
S5        & 05 46 11.6 & -00 06 28 & 23.74$\pm$0.27   & 21.00$\pm$0.04 & 18.96$\pm$0.03 & 17.22$\pm$0.02 \\ 
S6        & 05 46 09.6 & -00 03 31 & 23.26$\pm$0.18   & 20.35$\pm$0.02 & 18.33$\pm$0.02 & 16.85$\pm$0.02 \\
\enddata

\tablenotetext{a}{SDSS photometry taken from the galactic plane "Orion" release data (Finkbeiner et al. 2004) 
dataset 'calibImage-259-r3-0534' from 1998 Nov 17 (MJD~51134). Errors are from quoted inverse variance 
measurements on the PSF fluxes.}

\end{deluxetable}
\end{center}
\clearpage
%############################################################################################
% Table 3
\begin{center}
\begin{deluxetable}{ccccccccc}
\tabletypesize{\scriptsize}
\tablecaption{V1647~Ori Optical Photometry\label{photom}}
\tablewidth{0pc}
\rotate
\tablehead{
\colhead{MJD\tablenotemark{a}} & \colhead{g'$\pm$err(dg')} & \colhead{r'$\pm$err(dr')} &  
\colhead{i'$\pm$err(di')} & \colhead{z'$\pm$err(zg')} & \colhead{g'-r'}  & \colhead{r'-i'} & 
\colhead{i'-z'} & \colhead{Q$_{riz}$\tablenotemark{b}} \\
\colhead{}                     & \colhead{(mags)}          & \colhead{(mags)}          & 
\colhead{(mags)}          & \colhead{(mags)}          & \colhead{(mags)} & \colhead{(mags)} 
& \colhead{(mags)} & \colhead{(mags)} \\
}
\startdata
1135\tablenotemark{c} & --  & 23.04$\pm$0.22(6.34) & 20.81$\pm$0.05(5.91) & 18.80$\pm$0.04(4.41) & 1.70 & 2.23 & 2.01 & 0.25$\pm$0.23 \\
3049 & 20.05$\pm$0.15(0.00) & 17.70$\pm$0.09(0.00) & 15.90$\pm$0.08(0.00) & 14.39$\pm$0.06(0.00) & 2.35 & 1.80 & 1.51 & 0.31$\pm$0.20 \\
3074 &  --   --             & 17.91$\pm$0.09(0.21) &  --   --             &  --   --             & --   & --   & --   & \\
3251 & 20.06$\pm$0.15(0.01) & 17.81$\pm$0.09(0.11) & 15.93$\pm$0.08(0.03) & 14.39$\pm$0.06(0.00) & 2.25 & 1.88 & 1.54 & 0.36$\pm$0.20 \\
3284 & 20.27$\pm$0.15(0.21) & 18.03$\pm$0.09(0.33) & 16.20$\pm$0.08(0.30) & 14.66$\pm$0.06(0.27) & 2.24 & 1.83 & 1.54 & 0.31$\pm$0.20 \\
3322 & 20.28$\pm$0.15(0.23) & 18.02$\pm$0.09(0.32) & 16.19$\pm$0.08(0.29) & 14.65$\pm$0.06(0.26) & 2.26 & 1.83 & 1.54 & 0.31$\pm$0.20 \\
3351 & 20.75$\pm$0.15(0.70) & 18.38$\pm$0.09(0.68) & 16.52$\pm$0.08(0.62) & 15.06$\pm$0.06(0.67) & 2.37 & 1.86 & 1.46 & 0.42$\pm$0.20 \\
3378 & 20.19$\pm$0.15(0.14) & 17.89$\pm$0.09(0.19) & 16.10$\pm$0.08(0.20) & 14.60$\pm$0.06(0.21) & 2.30 & 1.79 & 1.50 & 0.31$\pm$0.20 \\
3612 & 21.06$\pm$0.15(1.01) & 18.69$\pm$0.09(0.99) & 16.86$\pm$0.08(0.96) & 15.37$\pm$0.06(0.98) & 2.37 & 1.83 & 1.49 & 0.36$\pm$0.20 \\
3634 & 21.27$\pm$0.15(1.22) & 18.87$\pm$0.09(1.17) & 17.00$\pm$0.08(1.10) & 15.54$\pm$0.06(1.15) & 2.40 & 1.87 & 1.46 & 0.43$\pm$0.20 \\
3656 & 21.58$\pm$0.15(1.53) & 19.26$\pm$0.09(1.56) & 17.37$\pm$0.08(1.47) & 15.90$\pm$0.06(1.51) & 2.32 & 1.89 & 1.47 & 0.44$\pm$0.20 \\
3692 & --   --              & 21.01$\pm$0.09(3.31) &  --   --             &  --   --             & --   & --   & --   & \\
3702 & 23.60$\pm$0.15(3.55) & 21.10$\pm$0.09(3.40) & 19.07$\pm$0.08(3.17) & 17.42$\pm$0.06(3.03) & 2.50 & 2.03 & 1.65 & 0.40$\pm$0.20 \\
3729 & 24.00$\pm$0.15(3.95) & 21.58$\pm$0.09(3.88) & 19.49$\pm$0.08(3.59) & 17.79$\pm$0.06(3.40) & 2.42 & 2.09 & 1.70 & 0.41$\pm$0.20 \\
3740 & 24.44$\pm$0.15(4.39) & 21.94$\pm$0.09(4.24) & 19.74$\pm$0.08(3.84) & 18.07$\pm$0.06(4.58) & 2.50 & 2.20 & 1.67 & 0.55$\pm$0.20 \\
3782 & --   --              & 22.89$\pm$0.09(5.19) & 20.69$\pm$0.08(4.79)\tablenotemark{d} &  --   --   & --   & --   & --   & \\
4090 & --   --              & 22.78$\pm$0.25(5.08)\tablenotemark{e} & -- --  &  --   --             & --   & --   & --   & \\
\enddata

\tablenotetext{a}{Modified Julian Date. 2450000+}
\tablenotetext{b}{Reddening invariant colors as defined by McGehee et al. (2004).}
\tablenotetext{c}{SDSS photometry taken from the galactic plane Orion release data (Finkbeiner et al. 2004) from MJD51134}
\tablenotetext{d}{i' photometry derived from r' and r'-i' color from MJD~3740.}
\tablenotetext{e}{R photometry from image taken at University of Hawaii 2.2m telescope.}
\end{deluxetable}
\end{center}
\clearpage
%############################################################################################
% Table 4
%\begin{center}
\begin{deluxetable}{rccccccccccccccccc}
\tabletypesize{\scriptsize}
%{\scriptsize}
\tablecaption{Identification of Spectral Features\tablenotemark{a}\label{lineids}}
\tablewidth{0pc}
\rotate
\tablehead{
\colhead{MJD} & 
\colhead{3049} & 
\colhead{3053} & 
\colhead{3074} & 
\colhead{3251} & 
\colhead{3257} & 
\colhead{3284} & 
\colhead{3322} & 
\colhead{3378} & 
\colhead{3612} & 
\colhead{3634} & 
\colhead{3656} & 
\colhead{3692} & 
\colhead{3701} & 
\colhead{3702} & 
\colhead{3740} & 
\colhead{3782} & 
\colhead{4152} \\
\colhead{Line (\AA)} & 
\colhead{} & 
\colhead{} & 
\colhead{} & 
\colhead{} & 
\colhead{} & 
\colhead{} & 
\colhead{} & 
\colhead{} & 
\colhead{} & 
\colhead{} & 
\colhead{} & 
\colhead{} & 
\colhead{} & 
\colhead{} & 
\colhead{} & 
\colhead{} & 
\colhead{}}

\startdata
YY                   & 04 & 04 & 04 & 04 & 04 & 04 & 04 & 05 & 05 & 05 & 05 & 05 & 05 & 06 & 06 & 06 & 07 \\
MM                   & 02 & 02 & 03 & 09 & 10 & 11 & 12 & 01 & 08 & 09 & 10 & 11 & 11 & 12 & 01 & 02 & 02 \\
DD                   & 14 & 18 & 10 & 03 & 06 & 13 & 12 & 08 & 30 & 25 & 13 & 19 & 27 & 25 & 05 & 16 & 21 \\
\hline \\
$[$Fe~II$]$~5226     & -- & -- & -- & E  & WE & WE & E  & WE & -- & -- & -- & -- & -- & -- & -- & -- & -- \\
He~I~5876            & -- & -- & A  & A  & WA & A  & WA & A  & WA & -- & -- & -- & -- & -- & -- & -- & -- \\
Na~D~5890+5896       & A  & AE & AE & AE & A  & A  & A  & A  & A  & AE & WA & -- & ?  & -- & -- & ?  & -- \\
Fe~I~6192            & E  & -- & WE & E  & E  & E  & E  & E  & E  & E  & WE & -- & -- & -- & -- & -- & -- \\
$[$O~I$]$~6300       & -- & E  & WE & E  & E  & E  & E  & E  & E  & E  & E  & E  & E  & E  & E  & E  & E  \\
Si~II~6347           & WA & A  & A  & A  & A  & A  & A  & A  & A  & A  & -- & -- & -- & -- & -- & -- & -- \\
$[$O~I$]$~6363       & -- & -- & E  & E  & E  & WE & E  & E  & WE & -- & E  & E  & E  & E  & E  & E  & -- \\
Si~II~6371           & WA & A  & A  & A  & A  & A  & A  & A  & A  & A  & WA & WA & -- & -- & -- & -- & -- \\
$[$Fe~II$]$~6432     & E  & -- & E  & E  & E  & E  & E  & E  & E  & WE & E  & WE & -- & -- & -- & -- & -- \\
Fe~I~6495            & -- & -- & E  & E  & E  & E  & E  & E  & E  & E  & E  & WE & E  & -- & -- & -- & -- \\
Fe~II~6517           & E  & -- & E  & E  & E  & E  & E  & E  & E  & E  & E  & -- & -- & E  & -- & -- & -- \\
H$\alpha$~6563       & P  & P  & P  & P  & P  & P  & P  & P  & P  & P  & P  & P  & P  & P  & P  & P  & P  \\
$[$S~II$]$~6713+6731 & -- & -- & E  & E  & E  & E  & E  & E  & E  & E  & E  & E  & E  & E  & E  & E  & E  \\
He~I~6678            & A  & -- & A  & A  & -- & WE & E  & E  & -- & E  & WE & -- & -- & -- & -- & -- & -- \\
He~I~7066            & O  & -- & A  & WA & ?  & A  & A  & WA & WA & WE & E  & WE & -- & -- & -- & -- & -- \\
$[$Fe~II$]$~7155     & O  & -- & E  & E  & E  & E  & E  & E  & E  & E  & E  & E  & E  & E  & E  & E  & -- \\
$[$Ca~II$]$~7261     & O  & -- & O  & ?  & E  & E  & WE & X  & WE & -- & -- & -- & -- & -- & -- & -- & -- \\
$[$Ca~II$]$~7291     & O  & -- & O  & E  & E  & E  & E  & X  & E  & E  & E  & E  & -- & -- & -- & -- & -- \\
$[$Ca~II$]$~7324     & O  & -- & O  & E  & E  & E  & E  & E  & E  & E  & WE & -- & -- & -- & -- & -- & -- \\
$[$Fe~II$]$~7388     & O  & -- & O  & E  & E  & E  & E  & E  & WE & WE & WE & E  & -- & -- & -- & -- & -- \\
Fe~II~7462           & O  & -- & O  & E  & E  & E  & E  & E  & E  & WE & X  & WE & -- & -- & -- & -- & -- \\
Fe~II~7711           & O  & O  & O  & E  & E  & E  & E  & E  & E  & E  & X  & -- & E  & E  & E  & -- & -- \\
K~I~7699             & O  & O  & O  & P  & P  & P  & P  & P  & P  & P  & P  & WP & -- & E  & E  & -- & A  \\
O~I~7773             & O  & O  & O  & A  & A  & A  & A  & A  & A  & A  & A  & ?  & -- & ?  & ?  & ?  & WE \\
Mg~II~7877           & O  & O  & O  & WA & WA & WA & WA & WA & WE & -- & -- & -- & -- & -- & ?  & -- & ?  \\
He~I~7878            & O  & O  & O  & A  & A  & A  & A  & WA & WE & -- & -- & -- & -- & -- & ?  & -- & -- \\
Mg~II~7896           & O  & O  & O  & A  & A  & A  & A  & WA & -- & WA & -- & -- & -- & -- & -- & -- & -- \\
\enddata

%$[$Fe~II$]$~7515     & O  & E  & O  & -- & -- & -- & WE & -- & -- & -- & WE & -- & -- & WE & WE & WE & -- \\
%$[$Fe~II$]$~5547     & -- & -- & -- & -- & -- & -- & -- & ba & -- & -- & -- & -- & -- & -- & -- & -- &  \\

\tablenotetext{a}{E=emission, A=absorption, WE=weak, P=P~Cygni profile, X=bad pixels at location of line, ?=inconclusive, --=no line evident, O=outside observed spectral range.}
%\tablenotetext{b}{Possibly effected by broad/red-shifted Na~D component.}
\end{deluxetable}
%\end{center}
\clearpage
%############################################################################################
% Table 4
\begin{deluxetable}{cccccccccc}
\tabletypesize{\scriptsize}
{\scriptsize}
\tablecaption{V1647 Ori H$\alpha$ Line Profiles\label{haprof}}
\tablewidth{0pc}
\tablehead{
\colhead{MJD\tablenotemark{a}} & 
\colhead{S:N\tablenotemark{b}} &
\colhead{W$_{\lambda}$\tablenotemark{c}} & 
\colhead{$\lambda_{peak}$\tablenotemark{d}} & 
\colhead{$\delta\lambda_{peak}$($\delta$v$_{peak}$)\tablenotemark{e}} & 
\colhead{FWZI\tablenotemark{f}} & 
\colhead{FW2.5\%\tablenotemark{g}} & 
\colhead{FW10\%\tablenotemark{h}} & 
\colhead{$\Delta$v\tablenotemark{i} } & 
\colhead{v$_{char}$\tablenotemark{j} } \\  
\colhead{} & 
\colhead{} & 
\colhead{(\AA)} & 
\colhead{(\AA)} & 
\colhead{(\AA(km~s$^{-1}$))} & 
\colhead{(km~s$^{-1}$)} & 
\colhead{(km~s$^{-1}$)} & 
\colhead{(km~s$^{-1}$)} & 
\colhead{(km~s$^{-1}$)} & 
\colhead{(km~s$^{-1}$)}}

\startdata
3049 & 252(34) & --31.4$\pm$2.0  & 6563.2$\pm$0.1 & +0.4(18)    & 1392$\pm$30  & 968$\pm$30  & 534$\pm$30   & 552$\pm$30  &  516$\pm$20 \\
3074 & 400(39) & --50.9$\pm$2.0  & 6563.0$\pm$0.1 & +0.2(9)     & 1384$\pm$30  & 1030$\pm$30 & 562$\pm$30   & 270$\pm$30  &  294$\pm$20 \\
3257 & 282(42) & --24.1$\pm$2.0  & 6562.3$\pm$0.1 & +0.5(23)    &  958$\pm$30  & 664$\pm$30  & 460$\pm$30   & 308$\pm$30  &  230$\pm$20 \\
3273 & 352(10) & --18.0$\pm$1.0  & 6563.3$\pm$0.1 & +0.5(24)    &  960$\pm$10  & 592$\pm$30  & 440$\pm$10   & 255$\pm$20  &  210$\pm$10\tablenotemark{k} \\
3284 & 278(36) & --29.5$\pm$2.0  & 6562.5$\pm$0.1 & --0.3(--14) & 1208$\pm$30  & 822$\pm$30  & 564$\pm$30   & 273$\pm$30  &  220$\pm$20 \\
3322 & 261(41) & --22.1$\pm$2.0  & 6562.7$\pm$0.1 & --0.1(--5)  & 1002$\pm$30  & 660$\pm$30  & 450$\pm$30   & 332$\pm$30  &  208$\pm$20 \\
3351 & 212(29) & --29.2$\pm$2.0  & 6562.8$\pm$0.1 & 0.0(0)      & 1056$\pm$30  & 778$\pm$30  & 526$\pm$30   & 248$\pm$30  &  213$\pm$20 \\
3378 & 361(49) & --27.1$\pm$2.0  & 6563.2$\pm$0.1 & +0.4(18)    & 1180$\pm$40  & 856$\pm$30  & 556$\pm$40   & 238$\pm$30  &  167$\pm$20 \\
3612 & 160(23) & --21.2$\pm$2.0  & 6563.0$\pm$0.1 & +0.2(9)     &  840$\pm$40  & 740$\pm$30  & 476$\pm$40   & 158$\pm$30  &  150$\pm$20 \\
3638 & 108(13) & --29.2$\pm$2.0  & 6562.5$\pm$0.1 & --0.3(--14) &  930$\pm$50  & 714$\pm$30  & 464$\pm$50   & 112$\pm$30  &  144$\pm$20 \\
3656 & 141(17) & --30.2$\pm$2.0  & 6563.4$\pm$0.1 & +0.6(27)    & 1190$\pm$50  & 864$\pm$30  & 512$\pm$50   &  95$\pm$30  &  142$\pm$20 \\
3692 &  38(4)  & --37.2$\pm$2.0  & 6563.6$\pm$0.1 & +0.8(37)    &  970$\pm$80  & 858$\pm$30  & 518$\pm$80   & 109$\pm$60  &  127$\pm$50 \\
3701 &  29(2)  & --64.2$\pm$2.0  & 6563.4$\pm$0.1 & +0.6(27)    &  688$\pm$80  & 604$\pm$30  & 458$\pm$80   &  98$\pm$60  &  108$\pm$50 \\
3729 &  40(3)  & --64.9$\pm$2.0  & 6563.7$\pm$0.1 & +0.9(41)    &  876$\pm$100 & 704$\pm$30  & 536$\pm$100  & $<$100      &  -- \tablenotemark{l} \\
3740 &  39(3)  & --66.6$\pm$2.0  & 6563.9$\pm$0.1 & +1.1(50)    &  868$\pm$100 & 686$\pm$30  & 598$\pm$100  & $<$100      &  -- \\
3782 &  29(1)  & --98.7$\pm$2.0  & 6564.4$\pm$0.1 & +1.6(73)    &  770$\pm$100 & 730$\pm$30  & 632$\pm$100  & $<$100      &  -- \\
4152\tablenotemark{m} &  15(1)  & --136.2$\pm$2.0 & 6565.4$\pm$0.1 & +2.6(119)   &  822$\pm$100 & 762$\pm$30  & 756$\pm$100  & $<$100      &  -- \\
%\multicolumn{8}{c}{ } \\
%4709 & --38.2$\pm$2.0  & 6562.7$\pm$0.1 & --0.1 & 1640$\pm$100 & 1024$\pm$100 & 558$\pm$20  &  651$\pm$20 \\
\enddata

\tablenotetext{a}{Modified Julian Date. 2450000+}

\tablenotetext{b}{Signal to noise on the H$\alpha$ emission peak and, in
  parentheses, in the adjacent continuum.}

\tablenotetext{c}{Equivalent width of emission component calculated
  from the continuum levels on the blue and red edges of the main
  H$\alpha$ emission peak. The uncertainties are estimated from the
  spread of repeat measurements.}

\tablenotetext{d}{The wavelength of peak H$\alpha$ emission,
  $\lambda_{peak}$}

\tablenotetext{e}{Offset of peak H$\alpha$ emission from 6562.8~$\AA$
  in $\AA$ and, in parentheses, km~s$^{-1}$.}

\tablenotetext{f}{The Full-Width Zero Intensity of the H$\alpha$
  emission in km~s$^{-1}$. The velocity offset to the continuum on the
  red side of 6562.8~$\AA$ was measured then doubled to obtain the
  FWZI value. The uncertainties are estimated from the spread of
  repeat measurements.}

\tablenotetext{g}{The Full-Width Intensity of the H$\alpha$ emission
  in km~s$^{-1}$ measured to a signal of I$_{max}$/40 (2.5\%), where
  I$_{max}$ is the peak line intensity.  The width is measured on the
  red side of 6562.8~$\AA$ and was doubled to obtain the FW2.5\%
  value. The uncertainties are estimated from the spread of repeat
  measurements.}

\tablenotetext{h}{The Full-Width Intensity of the H$\alpha$ emission
  in km~s$^{-1}$ measured to a signal of 10\% of the peak intensity,
  I$_{max}$.  The width is measured on the red side of 6562.8~$\AA$
  and was doubled to obtain the FW10\% value.  The uncertainties are
  estimated from the spread of repeat measurements.}

\tablenotetext{i}{The width of the H$\alpha$ absorption feature in
  km~s$^{-1}$.  The uncertainties are estimated from the spread of
  repeat measurements.}

\tablenotetext{j}{The velocity offset (in km~s$^{-1}$) of the deepest
  absorption from 6562.8~$\AA$, the rest wavelength of H$\alpha$. The
  uncertainties are estimated from the spread of repeat measurements.}

\tablenotetext{k}{Keck/HIRES spectrum. Profile not shown in
  Figs.~\ref{haplot1} and \ref{haplot2}}

\tablenotetext{l}{Emission and absorption to weak to accurately
  measure at signal to noise present.}

\tablenotetext{m}{Data from Aspin et al. (2009).}

\end{deluxetable}
\clearpage

%############################################################################################
% Figure Idplot
%############################################################################################
% Figure Photplot
%############################################################################################
% Figure sestar
%############################################################################################
% Figure Colplot 
%############################################################################################
% Figure Ccplot
%############################################################################################
% Figure Colq
%###########################################################################################
% Figure Icplot
%############################################################################################
% Figure Rplot
%############################################################################################
% Figure rkplot
%############################################################################################
% Figure Fullplot
%############################################################################################
% Figure Oct04 plot
%############################################################################################
\clearpage
% Figure Na D plot
%############################################################################################
\clearpage
% Figure [SII] plots
%############################################################################################
\clearpage
% Figure OI 6300A plots
%############################################################################################
\clearpage
% Figure OI 6300A plots
%############################################################################################
\clearpage
% Figure Fe I and II plots
%############################################################################################
\clearpage
% Figure CaII plots
%############################################################################################
\clearpage
% Figure HIRES Ha fit plots
%############################################################################################
\clearpage
% Figure HIRES CaII plots
%############################################################################################
\clearpage
% Figure Ha profiles maxmin
%############################################################################################
% Figure Ha profiles absorption 
%############################################################################################
% Figure Ha profile variations 
%############################################################################################

\end{document}